\pdfoutput=1

\documentclass[11pt,twoside,a4paper,cmspaper,final,collab]{cms-tdr}
\def\svnVersion{446979P}\def\cmsCernNoTag{CERN-EP-2017-243}\def\cmsCernDate{\today}\def\cmsMessage{Published in the Journal of High Energy Physics as \href{http://dx.doi.org/10.1007/JHEP02(2018)067}{\doi{10.1007/JHEP02(2018)067.}}}

\begin{document}\cmsNoteHeader{SUS-16-041}

\hyphenation{had-ron-i-za-tion}
\hyphenation{cal-or-i-me-ter}
\hyphenation{de-vices}
\RCS$HeadURL: svn+ssh://svn.cern.ch/reps/tdr2/papers/SUS-16-041/trunk/SUS-16-041.tex $
\RCS$Id: SUS-16-041.tex 442190 2018-01-22 11:04:07Z ikhvastu $
\newlength\cmsFigWidth
\ifthenelse{\boolean{cms@external}}{\setlength\cmsFigWidth{0.85\columnwidth}}{\setlength\cmsFigWidth{0.4\textwidth}}
\ifthenelse{\boolean{cms@external}}{\providecommand{\cmsLeft}{top\xspace}}{\providecommand{\cmsLeft}{left\xspace}}
\ifthenelse{\boolean{cms@external}}{\providecommand{\cmsRight}{bottom\xspace}}{\providecommand{\cmsRight}{right\xspace}}
\newcommand{\x}{\ensuremath{\phantom{0}}}
\newcommand{\y}{\ensuremath{\phantom{.}}}

\newcommand{\MT}{\ensuremath{M_{\text{T}}}\xspace}
\newcommand{\MTmin}{\ensuremath{M_{\text{T}}^{\text{min}}}\xspace}
\newcommand{\Njets}{\ensuremath{N_{\text{jets}}}\xspace}
\newcommand{\Nbjets}{\ensuremath{N_{\text{b jets}}}\xspace}
\newcommand{\ptrel}{\ensuremath{{p}_{\text{T}}^{\text{rel}}}\xspace}
\newcommand{\mll}{\ensuremath{\mathrm{m}_{\ell\ell}}\xspace}
\newcommand{\WJ}{\ensuremath{\PW + \mbox{jets}}\xspace}
\newcommand{\ZJ}{\ensuremath{\PZ + \mbox{jets}}\xspace}
\newcommand{\WZ}{\ensuremath{\PW\PZ}\xspace}
\newcommand{\ttZ}{\ensuremath{\ttbar\PZ}\xspace}
\newcommand{\ttW}{\ensuremath{\ttbar\PW}\xspace}
\newcommand{\ttH}{\ensuremath{\ttbar\mbox{H}}\xspace}
\newcommand{\ttX}{\ensuremath{\ttbar\textrm{X}}\xspace}
\newcommand{\qqWW}{\ensuremath{\PW^{\pm}\PW^{\pm}}\xspace}

\newcommand{\pp}{\ensuremath{\text{pp}}\xspace}

\newcommand{\gluino}{\ensuremath{\widetilde{\text{g}}}\xspace}
\newcommand{\susytop}{\ensuremath{\widetilde{\text{t}}}\xspace}
\newcommand{\sbottom}{\ensuremath{\widetilde{\text{b}}}\xspace}
\newcommand{\lsp}{\ensuremath{\widetilde{\chi}^0_1}\xspace}
\newcommand{\chiplus}{\ensuremath{\widetilde{\chi}^-}\xspace}
\newcommand{\chimimus}{\ensuremath{\widetilde{\chi}^+}\xspace}
\newcommand{\chiplmin}{\ensuremath{\widetilde{\chi}^\pm}\xspace}
\newcommand{\chitz}{\ensuremath{\widetilde{\chi}^0_2}\xspace}

\newcommand{\stoptwo}{\ensuremath{\widetilde{\text{t}}_2}\xspace}
\newcommand{\stopone}{\ensuremath{\widetilde{\text{t}}_1}}

\newcommand{\ptRatio}{\ensuremath{\pt^\text{ratio}}\xspace}
\newcommand{\ptRel}{\ensuremath{\pt^\text{rel}}\xspace}
\newcommand{\relIso}{\ensuremath{I_\text{rel}}\xspace}
\newcommand{\miniIso}{\ensuremath{I_\text{mini}}\xspace}
\newcommand{\multiIso}{\ensuremath{I_\text{multi}}\xspace}
\providecommand{\NA}{\ensuremath{\text{---}}}

\cmsNoteHeader{SUS-16-041}
\title{Search for supersymmetry in events with at least three electrons or muons, jets, and missing transverse momentum in proton-proton collisions at $\sqrt{s}=13\TeV$}

\date{\today}

\abstract{
A search for new physics is carried out in events with at least three electrons or muons in any combination, jets, and missing transverse momentum. Results are based on the sample of proton-proton collision data produced by the LHC at a center-of-mass energy of 13\TeV and collected by the CMS experiment in 2016. The data sample analyzed corresponds to an integrated luminosity of 35.9\fbinv. Events are classified according to the number of b jets, missing transverse momentum, hadronic transverse momentum, and the invariant mass of same-flavor dilepton pairs with opposite charge. No significant excess above the expected standard model background is observed. Exclusion limits at 95\% confidence level are computed for four different supersymmetric simplified models with pair production of gluinos or third-generation squarks. In the model with gluino pair production, with subsequent decays into a top quark-antiquark pair and a neutralino, gluinos with masses smaller than 1610\GeV are excluded for a massless lightest supersymmetric particle. In the case of bottom squark pair production, the bottom squark masses are excluded up to 840\GeV for charginos lighter than 200\GeV. For a simplified model of heavy top squark pair production, the $\mathrm{\widetilde{\text{t}}_2}$ mass is excluded up to 720, 780, or 710\GeV for models with an exclusive $\mathrm{\widetilde{\text{t}}_2}\to\mathrm{\widetilde{\text{t}}_1}\mathrm{H}$ decay, an exclusive $\mathrm{\widetilde{\text{t}}_2}\to\mathrm{\widetilde{\text{t}}_1}\mathrm{Z}$ decay, or an equally probable mix of those two decays. In order to provide a simplified version of the analysis for easier interpretation, a small set of aggregate signal regions also has been defined, providing a compromise between simplicity and analysis sensitivity.
}

\hypersetup{%
pdfauthor={CMS Collaboration},%
pdftitle={Search for supersymmetry in events with at least three electrons or muons, jets, and missing transverse momentum in proton-proton collisions at sqrt(s) = 13 TeV},%
pdfsubject={CMS},%
pdfkeywords={CMS, physics, SUSY, leptonic}}

\maketitle

\section{Introduction}
\label{sec:Introduction}

Many different theories beyond the standard model
(BSM) predict
processes leading to events containing multiple electrons and/or muons~\cite{Eboli:1998yi,Craig:2012vj,Craig:2016ygr,Craig:2012pu,Chen:2013xpa}.
The
background from standard model (SM) processes forging such a final
state is small and dominated by multiboson production, which is well
understood theoretically~\cite{Lazopoulos:2008de,Kardos:2011na,Campbell:2012dh,Campbell:2013yla,Kulesza:2015vda,Broggio:2015lya,Grazzini:2017ckn,Cascioli:2014yka,Caola:2015psa,Campbell:2016ivq,Binoth:2008kt,Nhung:2013jta,Yong-Bai:2015xna,Yong-Bai:2016sal,Hong:2016aek} and well reconstructed experimentally~\cite{Aaboud:2016xve,Aaboud:2016yus,Khachatryan:2016tgp,Aad:2015zqe,Khachatryan:2016txa}.
The search in this paper is designed to have broad sensitivity to a
variety of BSM models by examining the event yields as a function of
several kinematic quantities.

This paper describes the methods and results of a search for new
physics in final states with three or more electrons or muons in any combination accompanied by
jets and missing transverse momentum. A sample of proton-proton ($\Pp\Pp$) collision data, corresponding to an integrated luminosity of 35.9\fbinv and collected by the CMS
detector at the CERN LHC at a center-of-mass energy of 13\,TeV throughout 2016, is used.
Results of this analysis are interpreted in the context of
supersymmetric (SUSY) models~\cite{Ramond:1971gb,Golfand:1971iw,Neveu:1971rx,Volkov:1972jx,Wess:1973kz,Wess:1974tw,Fayet:1974pd,Nilles:1983ge,Martin:1997ns}.
Supersymmetry is an extension of the SM that predicts a SUSY partner for every SM particle by introducing a new symmetry between
bosons and fermions. It can potentially provide
solutions to
questions left open by the SM, such as the hierarchy problem and the
nature of dark matter. More specifically, models in which
$R$-parity~\cite{Wess:1974tw} is conserved, whereby SUSY
particles are produced only in pairs, can include a dark matter candidate
in the form of a stable and undetectable lightest SUSY
particle (LSP). In the models considered in this paper, the LSP is
assumed to be the lightest neutralino (a mixture of the superpartners of the Higgs and Z bosons, and of the photon).

\begin{figure}[htbp]
\begin{center}
\includegraphics[width=0.32\textwidth]{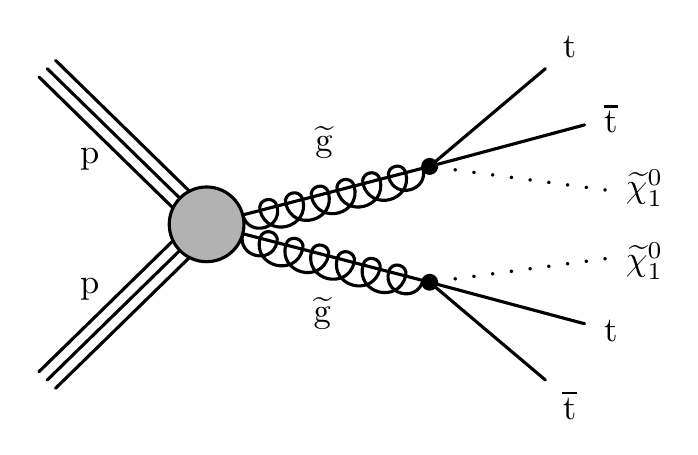}  \hfil
\includegraphics[width=0.32\textwidth]{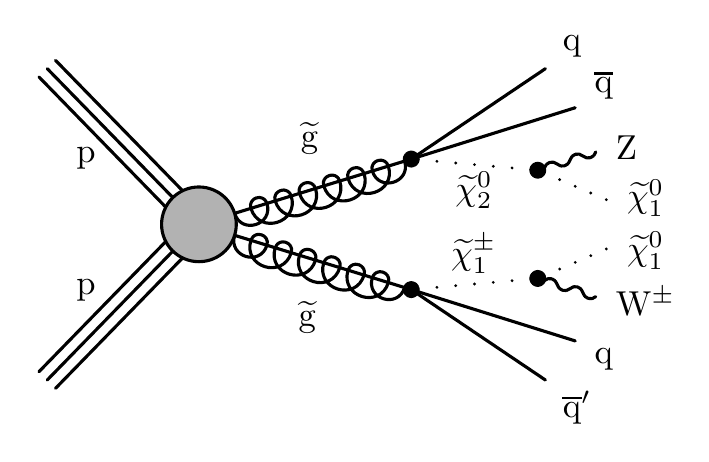}\\
\includegraphics[width=0.32\textwidth]{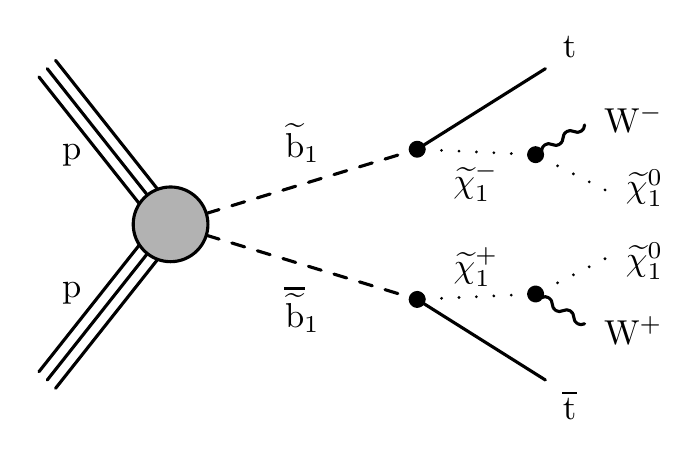}  \hfil
\includegraphics[width=0.32\textwidth]{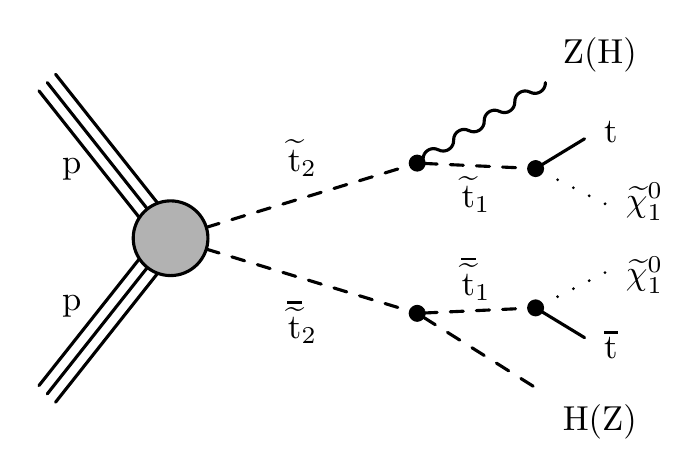}
\end{center}
\caption{\label{fig:diagrams} Diagrams for models with gluino pair production leading to four top quarks, T1tttt (upper left), or four quarks and two
vector bosons, T5qqqqVV (upper right) in the final state, in both cases accompanied by two LSPs. Models of bottom, T6ttWW, and top squark, T6ttHZ, pair production lead to two top quarks, two LSPs and either two W bosons (lower left) or two neutral bosons as SM Higgs (H) and/or Z bosons (lower right).}
\end{figure}

The reference models for this analysis are simplified model spectra
(SMS)~\cite{Alves:2011wf}. Examples for SUSY processes that can give
rise to multilepton final states are shown in Fig.~\ref{fig:diagrams}.
Throughout this paper lepton refers to an electron or a muon.
The models under consideration in this analysis
feature the pair production of gluinos, $\gluino$, or third generation squarks, $\sbottom_{1}$ or $\stoptwo$, superpartners of
gluons and third generation quarks, respectively, for a wide spectrum of possible
masses.
A typical process predicted by SUSY models consists of gluino pair production with each gluino decaying to a top quark pair, $\ttbar$, and an LSP, \lsp (Fig.~\ref{fig:diagrams}, upper left), or to a pair of quarks and a neutralino, \chitz, or chargino, $\chiplmin_{1}$. The latter
would then decay into a Z or W boson, and an LSP
(Fig.~\ref{fig:diagrams}, upper right). The first model is referred to as T1tttt and the second one as T5qqqqVV throughout this paper. Other models feature bottom
squark, $\sbottom_{1}$, pair production, with subsequent cascade decays resulting in
top quarks, W bosons and LSPs (Fig.~\ref{fig:diagrams}, lower left)
or pair production of the heaviest of the two top squark states, \stoptwo,
with subsequent decays to top quarks, Higgs or Z bosons, and LSPs (Fig.~\ref{fig:diagrams}, lower right).
The latter process allows a challenging scenario to be probed in which the mass difference between
the lighter top squark, \stopone, and the neutralino, \lsp, is close to the mass of the top quark~\cite{Khachatryan:2014doa,Aaboud:2017ejf}. These two models are denoted as T6ttWW and T6ttHZ, respectively.
Through the decays of W, Z or Higgs bosons these processes can
result in several leptons. In addition to the presence of
multiple leptons, these models predict events with multiple jets and
missing transverse momentum, largely induced by the undetected LSPs.
The SUSY particles that are not directly included in the diagrams are
assumed to be too heavy to be accessible at the LHC. Therefore, the only free parameters in these models are the mass of the produced gluinos or squarks, the masses of the possible intermediate particles in the decay chain, like $\chitz$ or $\chiplmin_{1}$, and the mass of the $\lsp$.

Similar searches have been carried out by the ATLAS
and CMS
Collaborations using the 13\TeV dataset. With the data sample collected by the ATLAS experiment and corresponding to an integrated luminosity of 36.1\fbinv, gluinos with masses up to 1870\GeV can be excluded~\cite{Aaboud:2017dmy} assuming the model depicted in Fig.~\ref{fig:diagrams} (upper left).
A comparable search at the same center-of-mass energy with the CMS
detector in 2015, based on a data sample corresponding to an
integrated luminosity of 2.3\fbinv, excluded gluino masses
below 1175\GeV~\cite{SUS-16-003}. The current analysis improves upon
the one performed with the data collected in 2015 with a more advanced
strategy that exploits the transverse mass reconstructed with a lepton
and the missing transverse momentum vector. Taking into account that
approximately 15 times more data were collected in 2016, a new control
region dominated by events from the \ttZ process and a new
interpretation of the results based on a
T6ttHZ model also were added.

\section{The CMS detector}
\label{sec:cmsdetector}

The CMS detector features a superconducting solenoid with an internal diameter of 6\unit{m} that creates a magnetic field of 3.8\unit{T}. Inside the magnet volume are a silicon pixel and strip tracker, an electromagnetic calorimeter (ECAL) made of lead tungstate crystals, and a hadronic calorimeter  (HCAL)  made of  brass and  scintillator material, each  composed of a barrel  and  two  endcap sections. Forward calorimeters extend the pseudorapidity ($\eta$) coverage for the HCAL. In the barrel section of the ECAL, an energy resolution of about 1\% is achieved for unconverted or late-converting photons in the tens of GeV energy range. The remaining barrel photons have a resolution of about 1.3\% up to $\abs{\eta} = 1$, rising to about 2.5\% at $\abs{\eta} = 1.4$. In the endcaps, the resolution of unconverted or late-converting photons is about 2.5\%, while the remaining endcap photons have a resolution between 3 and 4\%~\cite{CMS:EGM-14-001}. When combining information from the entire detector, the jet energy resolution amounts typically to 15\% at 10\GeV, 8\% at 100\GeV, and 4\% at 1\TeV, to be compared to about 40\%, 12\%, and 5\% obtained when the ECAL and HCAL calorimeters alone are used.
Muons are measured in the range $\abs{\eta} < 2.4$, with detection planes made using three technologies: drift tubes, cathode strip chambers, and resistive plate chambers. Matching muons to tracks measured in the silicon tracker results in a relative transverse momentum resolution for muons with $20 <\pt < 100\GeV$ of 1.3--2.0\% in the barrel and better than 6\% in the endcaps, The \pt resolution in the barrel is better than 10\% for muons with \pt up to 1\TeV~\cite{Chatrchyan:2012xi}. The first level of the CMS trigger system~\cite{CMS-PAPERS-TRG-12-001}, composed of specialized hardware processors, uses information from the calorimeters and muon detectors to select the most interesting events in a fixed time interval of less than 4\unit{$\mu$s}. The high-level trigger (HLT) processor farm further decreases the event rate from approximately 100\unit{kHz} to around 1\unit{kHz},  before the storage of the data. A more detailed description of the CMS detector, together with a definition of the coordinate system used and the relevant kinematic variables, can be found in \cite{Chatrchyan:2008zzk}.

\section{Event selection criteria and Monte Carlo simulation}
\label{sec:selection}

Events are reconstructed using the particle flow, PF,
algorithm~\cite{Sirunyan:2017ulk}, which reconstructs and
identifies each individual particle with an optimized combination of
information from the various elements of the CMS detector. The objects
identified as particles by this algorithm are commonly referred to as
PF candidates. Jets are clustered from PF candidates using the
anti-$\kt$ algorithm~\cite{Cacciari:2008gp,Cacciari:2011ma} with a
distance parameter of 0.4. Only jets with transverse momentum (\pt)
larger than $30\GeV$ falling within $|\eta|<2.4$ are considered. To
avoid double counting, the closest matching jets to leptons are not
considered if they are separated from the lepton by less than 0.4 in
$\Delta R \equiv \sqrt{\smash[b]{(\Delta\eta)^2 +(\Delta\phi)^2}}$. Here
$\Delta\eta$ and $\Delta\phi$ are the differences in $\eta$ and
azimuthal angle ($\phi$, in radians) between the considered lepton and a given jet. Additional criteria are applied to reject events containing noise and mismeasured jets. Jet energy scale (JES) corrections are applied to correct simulated jets for residual differences with data~\cite{Chatrchyan:2011ds, Khachatryan:2016kdb}.

The combined secondary vertex algorithm CSVv2~\cite{CMS-BTV-12-001,BTV-16-002} is used to assess the likelihood that a jet originates from a bottom quark. The tagging efficiency for true b flavor jets is typically 70\% and the misidentification probabilities are 10\% and 1\% for c quark and light-flavor jets, respectively. Jets with $ \pt > 25\GeV$ and within $|\eta|<2.4$ are considered for b tagging. Another variable related to jets that is used throughout this analysis is the
scalar sum of the transverse momenta of all jets,
defined as $\HT = \sum_{\text{jets}} \pt$, where jets have
$\pt>30\GeV$. The missing transverse momentum \ptmiss is defined as
the magnitude of \ptvecmiss, the negative vector sum of the transverse
momenta all PF candidates reconstructed in an
event~\cite{JME-13-003,JME-16-004}.

Electron candidates are reconstructed using tracking and ECAL information, by combining the clusters of energy deposits in the ECAL with Gaussian sum filter tracks~\cite{Khachatryan:2015hwa}.
The electron identification is performed using a multivariate
discriminant built with shower shape variables, track-cluster matching variables, and track quality
variables. The algorithm is optimized to select electrons from the decay of $\PW$ and $\PZ$ bosons with a 90\% efficiency while rejecting electron candidates originating from jets. To reject electrons originating from photon conversions inside the detector, electrons are required to have all possible measurements in the innermost tracker layers and to be incompatible with any conversion-like secondary vertices. The identification of the muon is
performed using the quality of the
matching between the measurements of the
tracker and the muon system~\cite{Chatrchyan:2012xi}. The muon identification efficiency is at least 96\%, with
some variation depending on \pt and $\eta$.

The reconstructed vertex with the largest value of summed physics object
$\pt^2$ is taken to be the primary $\Pp\Pp$ interaction vertex. The physics
objects are the objects returned by a jet finding
algorithm~\cite{Cacciari:2008gp,Cacciari:2011ma} applied to all charged tracks
associated with the vertex, plus the corresponding associated missing transverse momentum.
Both muon and electron candidates are required to have a transverse
impact parameter smaller than 0.5\unit{mm} with respect to the primary vertex and a longitudinal impact parameter
smaller than 1\unit{mm}. In addition, a selection on the three-dimensional impact parameter
significance, defined as the value of impact parameter divided by its
uncertainty, is applied. This value has to be smaller than 4 for both
electrons and muons.

Additional information about the isolation of the lepton is necessary to discriminate between leptons originating from decays of heavy particles such as W and Z bosons (``prompt'' leptons) and those produced in hadron decays or jets misidentified as leptons (``nonprompt'' leptons). The lepton isolation criterion is constructed using three different variables.

The relative isolation, $\relIso$, is defined as the ratio of the
amount of energy measured in a cone around the lepton to the \pt of
the lepton, $\pt^\ell$, with a $\pt^\ell$-dependent radius~\cite{Khachatryan:2016kod}:
\begin{equation}
\Delta R \leq
\frac{10\GeV}{\min(\max(\pt^\ell, 50\GeV), 200\GeV)} .
\end{equation}

Requiring $\relIso$ below a given threshold ensures that the lepton is locally isolated, even in Lorentz-boosted topologies.

The second isolation variable is the ratio of the lepton \pt and that of the jet geometrically closest to the lepton: $\ptRatio~=~\pt^\ell/\pt^\text{jet}$. In most cases this is the jet containing the lepton. If no jet is found within a cone defined by $\Delta R < 0.4$, the ratio is set to 1. The use of \ptRatio provides a way to identify nonprompt low-\pt leptons originating from low-\pt b jets, which decay with a larger opening angle than the one used
in $\relIso$.

The last variable used in the isolation criteria of leptons is \ptRel, defined as the magnitude of the component of the lepton momentum perpendicular to the axis of the closest jet. The jet axis is obtained by subtracting the momentum vector of the lepton from that of the jet. If no matched jet is found around the lepton, the variable is set to 0. This variable allows the recovery of leptons from accidental overlap with jets in Lorentz-boosted topologies. For the calculation of \ptRel and the previously mentioned $\ptRatio$, jets with $\pt > 5\GeV$ and without any additional identification criteria are considered.

Using those three variables, a lepton is considered isolated if the following condition is fulfilled:
\begin{equation}
\relIso < I_1 \text{ AND } ( \ptRatio > I_2 \text{ OR } \ptRel > I_3 ) .
\end{equation}

The values of $I_1$, $I_2$, and $I_3$
depend on the flavor of the lepton; the probability to misidentify a
jet as a lepton is higher for electrons than for muons, so tighter
isolation values are used for the former.
For electrons (muons), the tight selection requirements are $I_1 =
0.12$ (0.16), $I_2 = 0.76$ (0.69), and $I_3 = 7.2$ (6.0)\GeV. The
isolation requirement for leptons to pass the loose working point of
the selection is significantly relaxed, only consisting of $\relIso <
0.4$.

Events used in this analysis are required to pass trigger selection criteria that target dilepton and multilepton events.
The following two sets of triggers are used in a logic OR configuration.
One set of triggers requires that the two leptons satisfy loose isolation criteria and that the highest-\pt (leading) lepton have $\pt > 23 \; (17)\GeV$
and the second highest-\pt (sub-leading) lepton have $\pt > 12 \; (8)\GeV$ for muons (electrons).
The second set of triggers places no requirements on the isolation, has a lower \pt threshold for both leptons
($\pt > 8\GeV$), and requires the \HT reconstructed in the trigger to
be greater than 300\GeV. With the thresholds on the \pt of the leptons and on the \HT applied, the efficiency per event is near $100\%$.

The selection requires the presence of at least three well-identified
leptons in the event. The leptons must satisfy \pt thresholds that depend on the lepton flavor and the amount of hadronic activity
in the event. For events with low hadronic activity ($\HT <
400\GeV$), the leading electron (muon) must satisfy $\pt> 25 \; (20)\GeV$ and
sub-leading electrons (muons) must satisfy $\pt> 15$ (10)\GeV.
In events with high hadronic activity ($\HT >
400\GeV$), the thresholds are relaxed to 15 (10)\GeV for the leading electrons (muons). The lowest-\pt (trailing) lepton must have
$\pt>10\GeV$ in all cases. Opposite-charge same-flavor lepton pairs
are required to have an invariant mass ($m_{\ell\ell}$) greater than
12\GeV to suppress Drell--Yan and quarkonium processes.

In order to estimate the contribution from SM processes with prompt leptons in the signal regions
and to calculate the predicted yields from new physics models, Monte Carlo (MC) simulations are used.
The \MGvATNLO\!v2.2.2 or v2.3.3 generator~\cite{MADGRAPH5} was used to simulate events for the \ttbar, $\PW \Pgg^{*}$ and $\cPqt \PW \PZ$ processes, at leading order (LO), and
for \ttZ, \ttW, $\cPqt\PZ \Pq$, $\cPqt\PH \Pq$, $\cPqt\PH \PW$, $\PW \PW \PZ$, $\PW \PZ \PZ$, $\PZ\PZ \PZ$, $\ttbar\Pgg$, and $\PZ \Pgg^*$ final states, at next-to-leading order (NLO) in perturbative quantum chromodynamics.
The NLO \POWHEG~v2~\cite{Alioli:2010xd} generator is exploited for the \ttH~\cite{Hartanto:2015uka} and diboson~\cite{Melia:2011tj,Nason:2013ydw} production.
The NNPDF3.0LO~\cite{Ball:2014uwa} parton distribution functions (PDFs) are used for the simulated samples generated at LO
and the NNPDF3.0NLO~\cite{Ball:2014uwa} PDFs for those generated at NLO.
Parton showering and hadronization are simulated using the \PYTHIA~v8.212 generator~\cite{Sjostrand:2007gs}
with the CUETP8M1 tune~\cite{Skands:2014pea,CMS-PAS-GEN-14-001}.
A double-counting of the partons generated with \MGvATNLO\!and those with \PYTHIA is removed using
the MLM~\cite{Alwall:2007fs} and the {\sc FxFx}~\cite{Frederix:2012ps} matching schemes, in the LO and NLO samples, respectively.
The CMS detector response is modeled using a \GEANTfour-based model~\cite{Geant}.
The simulated samples include additional simultaneous interactions per bunch crossing (pileup),
with distributions that are weighted to match the observed data.

Monte Carlo simulation of signal events used for interpretation of the final results is done with
the \MGvATNLO program at LO precision, allowing for up to two additional partons in the calculation of the matrix elements.
The SUSY particle decays, parton showering, and hadronization are simulated with \PYTHIA v8.212.
The detector response for signal events is simulated using a CMS fast-simulation package~\cite{Abdullin:2011zz}
that is validated with respect to the \GEANTfour-based model. All simulated events are processed with the same
reconstruction procedure as data. Cross sections for SUSY signal processes, calculated at NLO with
next-to-leading-logarithmic (NLL) resummation, were provided by
the LHC SUSY Cross Section Working Group~\cite{Beenakker:1996ch,Kulesza:2008jb,Kulesza:2009kq,Beenakker:2009ha,Beenakker:2011fu,Borschensky:2014cia}.

\section{Search strategy}
\label{sec:strategy}

A baseline selection is applied to the dataset containing events of interest: three or more electrons or muons, at least two jets ($\Njets \geq 2$), $\ptmiss \geq 50\GeV$, and $m_{\ell\ell} \geq 12\GeV$ for all opposite-charge, same-flavor lepton pairs. All these requirements are listed in Table~\ref{tab:preselection}. Two different regions are defined, based on whether or not an event contains an opposite-charge, same-flavor lepton pair with an invariant mass within the 15\GeV window around the $\PZ$ boson mass~\cite{PDG2016}. If such a lepton pair is found the event is categorized as "on-Z", otherwise "off-Z".

\begin{table}[tbh]
\centering
\topcaption{Summary of all requirements used in baseline selection criteria.}
\label{tab:preselection}
\begin{tabular}{lc} \hline
Number of selected leptons & ${\geq}3$ \\
\Njets & ${\geq}2$ \\
\ptmiss,\GeV & ${>}50$ (70 in low \Nbjets and low \HT category) \\
\mll,\GeV & ${>}12$ \\  \hline
\end{tabular}
\end{table}

Events are further categorized into signal regions, which are defined according to several event observables: $\Nbjets$, $\HT$, $\ptmiss$, \mll, as well as the transverse mass reconstructed with a lepton and the missing transverse momentum vector,
\begin{equation}
\MT = \sqrt{2\pt^{\ell}\ptmiss\left[1-\cos\left(\phi_\ell-\phi_{\ptvecmiss}\right)\right]} .
\end{equation}

If the event is categorized as on-Z, the \MT is calculated with the lepton that is not involved in the Z boson mass reconstruction, otherwise the lepton yielding the lowest \MT value (\MTmin) is used in the computation of this variable.

The classification of selected events based on the number of b jets
creates signal regions with high signal-to-background ratios for
events from different signal models. For example, the T1tttt model
features several b jets, which would be categorized into signal
regions that are almost free of the leptonic WZ background owing to
the b jet requirements. Including the 0 b jet signal regions keeps the
analysis sensitive to signatures without b jets, such as T5qqqqVV model.
Additionally, a categorization in $\HT$ and $\ptmiss$
is useful to distinguish between compressed and noncompressed SUSY
spectra, i.e.\ models with small or large mass differences between the
SUSY particles in the decay chain.

Table~\ref{tab:srdefinition} shows the definition of the signal
regions  (SRs) into which the events passing the baseline selection
are subdivided. There are 16 separate off-Z and 16 on-Z SRs. Each category is split, depending on the number of b
jets (0, 1 and 2), the value of \HT (greater or lower than 400\GeV),
and \ptmiss (greater or lower than 150\GeV). These SRs are denoted as SR 1-12.
Motivated by the low expected yield of events with
high b jet multiplicities,
one inclusive SR with $\ptmiss<300\GeV$ and $\HT<600\GeV$
has been defined for ${\geq}3$ b jets (SR 13), and additionally to this three SRs with
significant amounts of $\HT$ ($>$600\GeV, SRs 14, 15) or $\ptmiss$ ($>$300\GeV, SR 16) have
been introduced, since various noncompressed SUSY models yield very high values for these variables. These latter three regions are
inclusive in the number of b jets. All of the 0 b jet regions, as well
as three regions with high \HT and \ptmiss values, are further split
depending whether \MT is smaller (designated with the letter "a" after the region number) or greater (designated with "b") than 120\GeV, leading to a total of 23 regions for each of the off-Z and on-Z categories.
In the on-Z regions with 0 or 1 b jet and $60 < \HT < 400\GeV$, the \ptmiss lower bound is raised to
70\GeV to completely suppress the contribution from the Drell--Yan
process.

\begin{table}[tbh]
\centering
\topcaption{Summary of the signal region definitions. The minimum \ptmiss requirement is raised from 50 to 70\GeV only for the on-Z SR1 and SR5.
Signal regions that are further subdivided at \MT = 120\GeV are indicated with $\dagger$. The search regions are mirrored for on- and off-Z categories.}
\label{tab:srdefinition}
\resizebox{1.0\textwidth}{!}{
\begin{tabular}{c|ccc|c|c}
\hline
$\Njets$ & $\Nbjets$                    & $\HT$ [\GeVns{}]                        & $50(70) \leq \ptmiss< 150\GeV$        & $150 \leq \ptmiss< 300\GeV$         &  $\ptmiss\geq 300\GeV$       \\  \hline
\multirow{8}{*}{${\geq}2$}     &  \multirow{2}{*}{0 } & \x60--400  &  \phantom{$\dagger$}   SR1 $\dagger$       &    \phantom{$\dagger$}  SR2 $\dagger$        & \multirow{8}{*}{SR16 $\dagger$}         \\
&                        & 400--600 &     \phantom{$\dagger$}  SR3 $\dagger$       &   \phantom{$\dagger$}   SR4 $\dagger$        &                                                          \\
& \multirow{2}{*}{1}  & \x60--400  &          SR5       &     SR6       &                                                         \\
&                       & 400--600 &         SR7       &     SR8        &                                                         \\
& \multirow{2}{*}{2 } & \x60--400  &       SR9       &   \x  SR10        &                                                            \\
&                      & 400--600 &      \x   SR11       &   \x SR12       &                                                        \\  \cline{4-5}
&${\geq}3$         & \x60--600  &      \multicolumn{2}{c|}{ \x \x SR13 }   &                                                   \\  \cline{4-5}
& inclusive                 & ${\geq} 600$       &   \x  \phantom{$\dagger$}   SR14 $\dagger$ &  \x \phantom{$\dagger$} SR15 $\dagger$   &                                                    \\ \hline
\end{tabular}}
\end{table}

In order to provide a simplified version of the analysis for easier
interpretation, a small set of aggregate signal regions has been
defined, providing a compromise between simplicity and analysis
sensitivity. The definition of these so-called super signal regions
(SSR) is given in Table~\ref{tab:ssrdefinition}. The additional requirement \MT greater than 120 GeV was added to the SSRs with respect to the relevant SRs.

\begin{table}[tbh]
\centering
\topcaption{Definition of the aggregate super signal regions (SSRs). This simpler classification is proposed for
reinterpretations, depending on the presence of a Z boson candidate and the number of b jets, along with additional simultaneous requirements on \MT, \ptmiss, and \HT.}
\label{tab:ssrdefinition}
\renewcommand{\arraystretch}{1.1}
\begin{tabular}{r|c|c}
\hline
 & $\Nbjets\leq2$, \ \ $\MTmin\geq120\GeV$ & $\Nbjets\geq3$, \ \ $\MTmin\geq120\GeV$   \\
 & $\HT\geq200\GeV$, \ \ $\ptmiss\geq250\GeV$ & $\HT\geq60\GeV$, \ \ $\ptmiss\geq50\GeV$ \\  \cline{2-3}
off-Z & SSR1 & SSR2  \\
on-Z & SSR3 & SSR4  \\  \hline
\end{tabular}
\end{table}

\section{Background estimation}
\label{sec:backgrounds}

All backgrounds leading to the multilepton final states targeted by this analysis can be subdivided into the categories listed below.

{\bf Nonprompt leptons} are leptons from heavy-flavor decays, misidentified hadrons,
muons from light-meson decays in flight, or electrons from unidentified photon conversions.
In this analysis $\ttbar$ events can enter the signal regions if
nonprompt leptons are present in addition to the prompt leptons from
the W boson decays. Top quark pair production gives the largest contribution for regions with
low $\HT$ and $\ptmiss$ values, and therefore
predominately populates signal regions 1 and 5, with 0 and 1 b jet,
respectively. Apart from \ttbar, Drell--Yan events can enter the
baseline selection.
However, they are largely suppressed by the $\ptmiss > 50\GeV$
selection, and additional rejection is achieved by increasing the
\ptmiss requirement to 70\GeV for on-Z regions with low \HT and low
\ptmiss. Processes that yield only one prompt lepton in addition to
nonprompt ones, such as W+jets and various single top quark channels, are
effectively suppressed by the three-lepton requirement because of the
low probability that two nonprompt leptons satisfy the tight
identification and isolation requirements. Albeit small, this
contribution is nevertheless accounted for in our method to estimate
the background due to nonprompt leptons (see below).

{\bf Diboson production} can yield multilepton final states with up to
three prompt leptons (WZ or $\PW\gamma^{*}$) and up to four prompt leptons (ZZ or $\PZ\gamma^*$), rendering irreducible backgrounds for this analysis. For simplicity, in the following we refer to these backgrounds as WZ and ZZ, respectively. The WZ production has a sizable contribution in the on-Z events, especially in the SRs without b jets. The yields of
these backgrounds in the various SRs are estimated by means
of MC simulation, with the normalization factors derived from control regions in data.

{\bf Other rare SM processes} that can yield three or more leptons
are \ttW, \ttZ, and triboson production. We also include the
contribution from the SM Higgs boson produced in association with a
vector boson or a pair of top quarks in this category of backgrounds,
as well as processes that produce additional leptons from internal
conversions, which are events that contain a virtual photon that
decays to leptons. The internal conversion background components,
X+$\gamma$, are strongly suppressed by the $\ptmiss > 50\GeV$ and
$\Njets \geq 2$ requirements. The background events containing top
quark(s) in association with a W, Z or Higgs boson or another pair of
top quarks are denoted as \ttX, except for \ttZ which is separately
delineated. For the estimation of the latter process, the same
strategy as for the WZ is used. All other processes are grouped into
one category that is denoted as rare SM processes. The contribution
from these processes as well as \ttX are estimated from MC simulation.

The background contribution from nonprompt leptons is estimated using the tight-to-loose ratio method~\cite{Khachatryan:2016kod}.
In this method, the yield is estimated in an application region
that is similar to the signal region but which contains
at least one lepton that fails the tight identification and
isolation requirements but satisfies the loose requirements.
The events in this region are
weighted by $f / (1-f)$,
where the tight-to-loose ratio $f$ is the probability that a loosely identified lepton also satisfies the full set of requirements.
This ratio is measured as a function of lepton \pt and $\eta$ in a control sample of multijet events that is enriched in nonprompt leptons
(measurement region). In this region, we require exactly one lepton, satisfying the loose object selection,
and one recoiling jet with $\Delta R(\text{jet},\ell)>1.0$ and $\pt >
30\GeV$ in the event. To suppress
processes that can contribute prompt leptons from a W or Z boson decay, such as W(+jets), DY or \ttbar, we additionally require both $\ptmiss$ and $\MT$ to be below 20\GeV. The remaining contribution
from these processes within the measurement region is estimated from MC simulation and subsequently
subtracted from the data.

In order to reduce the dependence of the tight-to-loose ratio on the flavor composition of the jets
from which the nonprompt leptons originate, this ratio is parameterized as a function of a variable that correlates more strongly with the mother parton \pt than with the lepton \pt. This variable is calculated by correcting the lepton \pt as a function of the energy in the isolation cone around it. This definition leaves the \pt of the leptons satisfying the tight
isolation criteria unchanged and modifies the \pt of those failing these criteria so that it is a better proxy for the mother parton \pt and results in a smaller variation as a function of the mother parton \pt.
The flavor dependence, which is much more important for the case of electrons, is further reduced by adjusting the loose electron
selection to obtain similar $f$ values for nonprompt electrons that originate from light- or heavy-flavor jets.
As a result, the tight-to-loose ratio measured in a multijet sample leads to a good description of nonprompt background
originating from $\ttbar$ events, which in most of the SR are dominant in this category of background.

The tight-to-loose ratio method for estimating the nonprompt background is validated both in a closure test in simulation and in a data control region orthogonal to the baseline selection with minimal signal contamination.
This region is defined by the requirement of three leptons that satisfy the nominal identification,
isolation, and \pt selection, one or two jets, $30 < \ptmiss < 50\GeV$, and no dilepton pair with an invariant mass compatible with a Z boson.
With these selection criteria a purity in \ttbar of 80\% can be achieved. We find an agreement of the order of 20--30\% between the predicted and observed yields in this control region.

The \WZ process is one of the main backgrounds in the regions with 0 b jets, while \ttZ gives a significant contribution in categories enriched in b jets. As mentioned earlier, the contribution of these backgrounds is estimated from simulation, but their normalizations are obtained from a simultaneous fit using two control regions, designed so that each is highly enriched in one of the processes. The \WZ control region is defined by the requirement of three leptons satisfying the nominal identification and isolation selections. Two leptons have to form an opposite charge, same flavor pair with $| m_{\ell\ell} - m_{\PZ} | <  15\GeV $, the number of jets and b jets
has to be ${\leq}1$ and 0, respectively. The \ptmiss has to be in the range $30 < \ptmiss < 100\GeV$, and \MT is required to be at least 50\GeV to suppress contamination from the Drell--Yan process. The purity of the \WZ control region is 80\%. The orthogonal control region for \ttZ is defined similarly to that for \WZ, except for a requirement on the number of jets: three leptons satisfying the nominal identification and isolation selection are to be found, two of them forming an opposite charge, same flavor pair with $| m_{\ell\ell} - m_{\PZ} | <  15\GeV $, at least 3 jets, and $30 < \ptmiss < 50\GeV$.
Events are classified by the number of b
jets, and three bins are formed for the \ttZ CR: the 0 b jet category,
where the background is dominated by the \WZ and \ttbar processes, and
the 1 and $\geq$2 b jet categories, enriched in \ttZ. The overall
purity of the \ttZ process is 20\%, increasing to 50\% in the bins
with at least one b jet. These three bins, together with the \WZ
control region are used in a simultaneous fit to obtain the scale
factors for the normalization of the simulated samples. In the fit to
data, the normalization and relative population across all four bins of all the components are allowed
to vary according to experimental and theoretical uncertainties. For
the \WZ process the obtained scale factor is compatible with unity,
$1.01 \pm 0.07$, and no correction is applied to the simulation, while
for the \ttZ it is found to be $1.14 \pm 0.28$. Therefore the yields
from the MC \ttZ sample obtained in the baseline region are scaled by
a factor of 1.14.

\section{Systematic uncertainties}
\label{sec:systematics}

The uncertainties in the expected SM backgrounds and signal yields are categorized as experimental,
such as those related to the JES or the b tagging efficiency description in the simulation;
theoretical, such as the uncertainties in the considered cross sections;
statistical, related to the observed yield in control regions in data;
and as uncertainties in the background estimation methods relying on control regions in data.
These uncertainties and their effect on the predicted yields are described below and summarized in Table~\ref{tab:systSummary}.

\begin{table}[b]
\topcaption{ The effect of the systematic uncertainties on the event yields of the backgrounds and signal processes. }
\label{tab:systSummary}
\begin{center}
\resizebox{1.0\linewidth}{!}{
\begin{tabular}{l c c c}
\hline
Source               & Effect on the backgrounds [\%]    & Effect on signal  [\%]        \\
\hline
Integrated luminosity          & 2.5                          & 2.5 \\
\hline
JES     & 1--8                         & \x1--10        \\
b tag efficiency      & 1--8                         & \x1--10       \\
Pileup               & 1--5                         & 1--5 \\
\hline
Lepton efficiencies  & 9                            & 15       \\
HLT efficiencies     & 3                            & 3          \\
\hline
Nonprompt application region statistics  & \x10--100   &\NA    \\
Nonprompt extrapolation & 30       &\NA \\
WZ control region normalization  & 10                           & \NA \\
\ttZ control region normalization  & 25                   &\NA\\
\hline
Limited size of simulated samples     & \x\x1--100       & \x10--100  \\
ISR modeling                          &     \NA      & \x1--10 \\
Modeling of unclustered energy        &     \NA      & \x1--20 \\
\hline
Ren., fact. scales, cross section (\ttW, \ttH)  &   11--13 &\NA \\
Ren., fact. scales, acceptance (\ttW, \ttZ, \ttH, signal)    &   \x3--18 & \x3--18    \\
PDFs (\ttW, \ttZ, \ttH)              &    2--3   &\NA                     \\
Other rare backgrounds    & 50        &\NA   \\
\hline
\end{tabular}
}
\end{center}
\end{table}

One of the major experimental sources of uncertainty is the knowledge of the JES. This uncertainty affects all simulated background and signal events. For the data set used in this analysis, the uncertainties in the jet energy scale
vary from 1\% to 8\%, depending on the transverse momentum and pseudorapidity of
the jet. The impact of these uncertainties is assessed by shifting the jet
energy correction factors for each jet up and down by one standard deviation and recalculating all kinematic quantities.
The systematic uncertainties related to JES corrections are also
propagated to the \ptmiss calculation. The propagation of the variation of the JES results in a variation of 1--10\% in the predicted event yields in the various signal regions of this analysis.

A similar approach is used for the uncertainties associated with the corrections for the $\cPqb$ tagging efficiencies for light, charm and bottom
flavor jets, which are parameterized as a function of $\pt$ and $\eta$.
The variation of the scale factor correcting for the differences
between data and simulation is at a maximum of the order of 10\% per
jet, and leads to an overall effect in the range of 1--10\% depending
on the signal region and on the topology of the event. The inaccuracy
of the inelastic cross section value that affects the pile up rate gives
up to a 5\% effect. The sources of uncertainties explained here were also studied for
the signal samples, and their impact on the predicted signal yields in
every search region has been estimated following the same procedures.

Lepton identification and isolation scale factors have been measured
as a function of lepton \pt and $\eta$. They are applied to correct
for residual differences in lepton selection efficiencies between data
and simulation. The corresponding uncertainties are estimated to be
about 3\% per lepton for both flavors, and additionally 2\% per lepton is
assigned to the signal leptons due to the detector fast simulation. Assuming
100\% correlation between the uncertainties on the corrections for the
different leptons, a flat uncertainty of 9\% is taken into account for
the background, while 15\% is considered for the signal. The uncertainty
related to the HLT trigger efficiency is evaluated to amount to
3\%.

For the nonprompt and misidentified lepton background, several
systematic uncertainties are considered. The statistical uncertainty
from the application region, which is used to estimate this background
contribution, ranges from 10 to 100\%. The regions where these
uncertainties are large are generally regions where the overall
contribution from this background is small.
The uncertainty arising from the
electroweak background subtraction in the measurement region for the
tight-to-loose ratio is propagated from the uncertainty on
the scale factor obtained from the fit to the control regions.
In the case where no events are
observed in the application region, an upper limit of the background
expectation is used as determined from the upper limit at 68\% confidence
level (CL) multiplied by the most likely
tight-to-loose ratio value.

The systematic uncertainty related to the extrapolation from the
control regions to the signal regions for the nonprompt lepton background
is estimated to be 30\%. This value has been extracted from closure tests performed by applying the method described in Section~\ref{sec:backgrounds} to simulated samples containing nonprompt leptons. From the simultaneous fit in the control regions, the uncertainty in the normalization of the \WZ process is estimated to be 10\%, while a value of 25\% is found for \ttZ background.

The limited size of the generated MC samples represents an additional source
of uncertainty. For the backgrounds that are estimated from simulation, such as \ttW, \ttZ and \ttH, as well as for all the
signal processes, this statistical uncertainty is computed from the number of
MC events entering the signal regions and varies widely across the SRs.

For signal efficiency calculations additional uncertainties in the description of the initial-state radiation (ISR) are taken into account.
The modeling of ISR by the version of the \MGvATNLO\!generator used for signal events was compared against a data sample
of $\ttbar$ events in the dilepton final state.
The corresponding corrections range from 0.51 to 0.92, depending on the jet multiplicity.
These corrections are then applied on simulated SUSY events based on the number of ISR jets to improve upon
the \MGvATNLO modeling of the multiplicity of additional jets from
ISR. Half the magnitude of these ISR corrections is assigned as an additional systematic uncertainty, which can be as large as 10\%.

The uncertainty in potential differences between the modeling
of \ptmiss in data and the fast simulation arising from unclustered energy
in the CMS detector is evaluated by comparing the reconstructed \ptmiss with the \ptmiss obtained using generator-level information.
This uncertainty ranges up to 20\%.

Theoretical uncertainties include the uncertainty in the renormalization ($\mu_{\textrm R}$) and factorization ($\mu_{\textrm F}$) scales,
and in the knowledge of the PDFs. These uncertainties are evaluated for several processes, namely \ttW, \ttZ, and
\ttH, which are dominant backgrounds in several signal regions.
Both the changes in the acceptance and cross sections related to these effects are taken into
account and propagated to the final uncertainties.

For the study of the renormalization and factorization uncertainties, variations up and down by a factor of two
with respect to the nominal values of $\mu_{\textrm F}$ and $\mu_{\textrm R}$ are
evaluated. The maximum difference in the yields with respect to the nominal case is observed when both scales
are varied up and down simultaneously. The effect on the overall cross section is
found to be ${\sim}13\%$ for \ttW and ${\sim}11\%$ for \ttH backgrounds. The effect of the variations of $\mu_{\textrm F}$ and $\mu_{\textrm R}$ on the acceptance is taken as additional, uncorrelated uncertainty on the acceptance corresponding to different
signal regions. This effect is found to vary between 3\% and 18\% depending on the SR and the process.

The uncertainty related to the PDFs is estimated from the 100 NNPDF
3.0 replicas, computing the deviation with respect to the nominal
yield for each of them in every signal region (the cross section and
acceptance effect are considered together)~\cite{Ball:2014uwa}. The root-mean-square of
the variations is taken as the value of the systematic
uncertainty. Since no significant differences between
signal regions have been found, a flat uncertainty of 3\% (2\%) is considered for \ttW (\ttZ and \ttH) backgrounds.
This value also includes the effect of the strong coupling constant
variation,
$\alpha_{\rm S}(M_{\PZ})$, which is added in quadrature. An extra, conservative, flat uncertainty of 50\% is assigned to the yield of the
remaining rare processes, which are not well measured.

\section{Results}
\label{sec:results}

\begin{figure}[h]
\centering
\includegraphics[width=.32\textwidth]{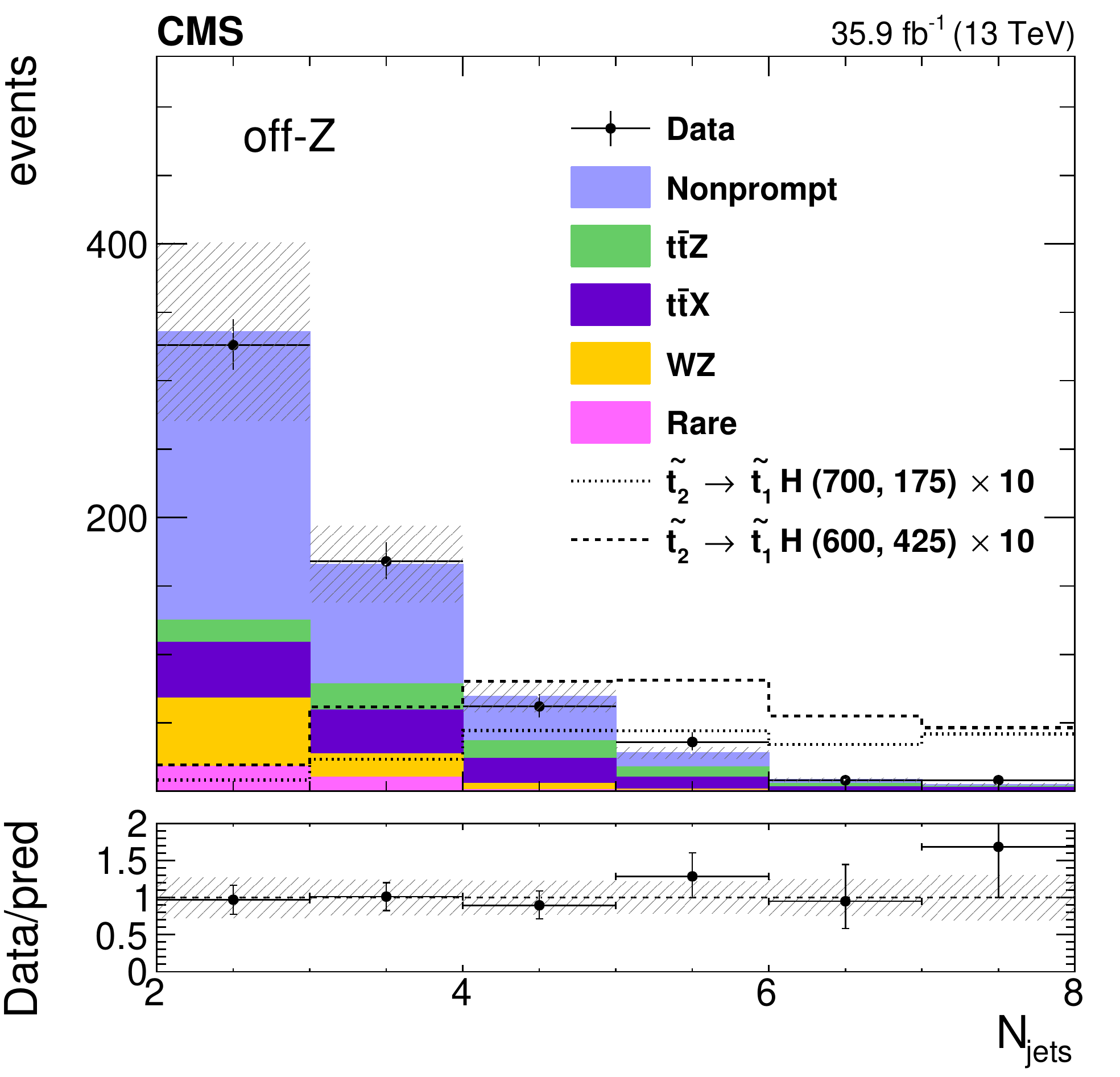}
\includegraphics[width=.32\textwidth]{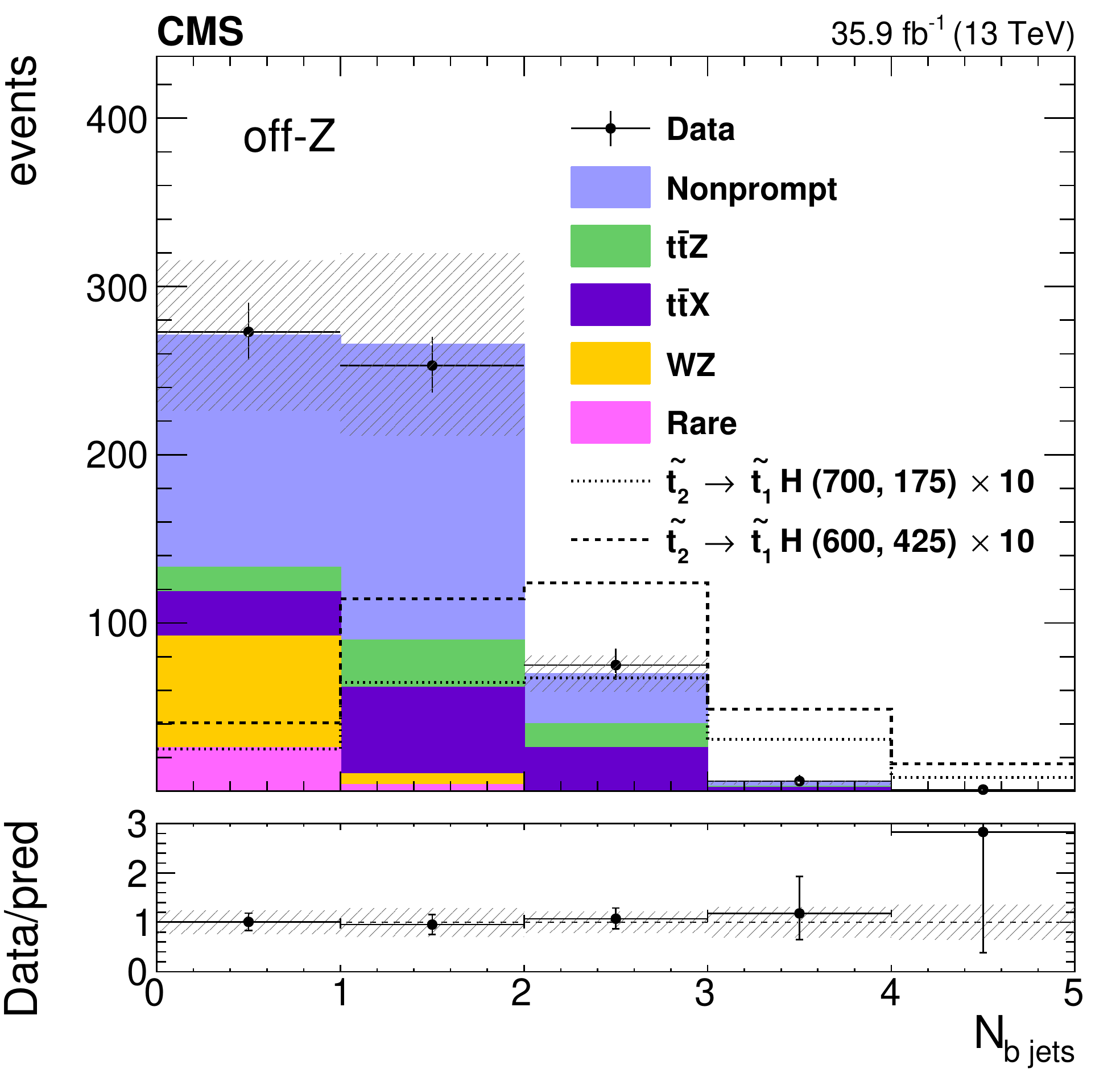}
\includegraphics[width=.32\textwidth]{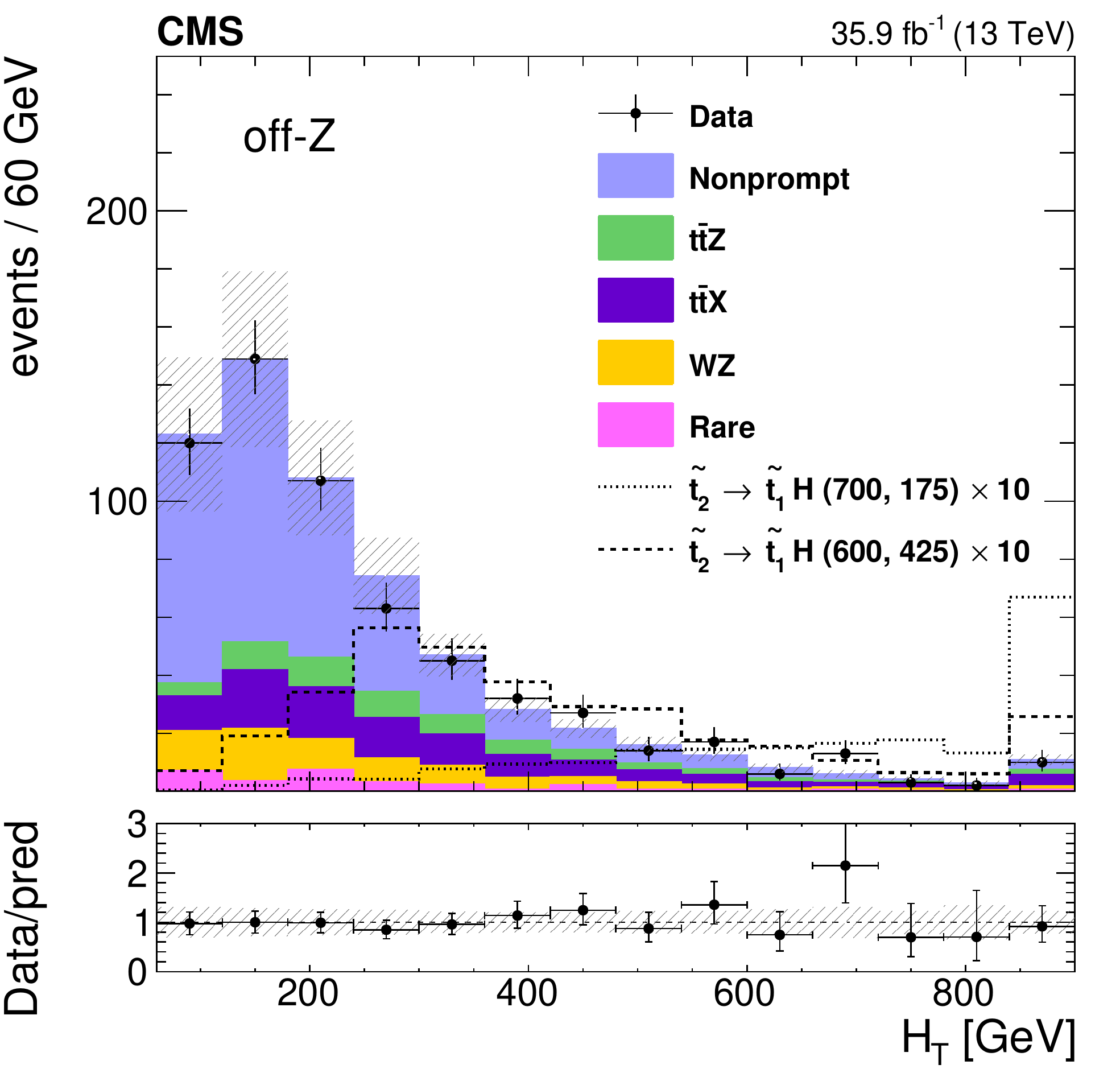}\\
\includegraphics[width=.32\textwidth]{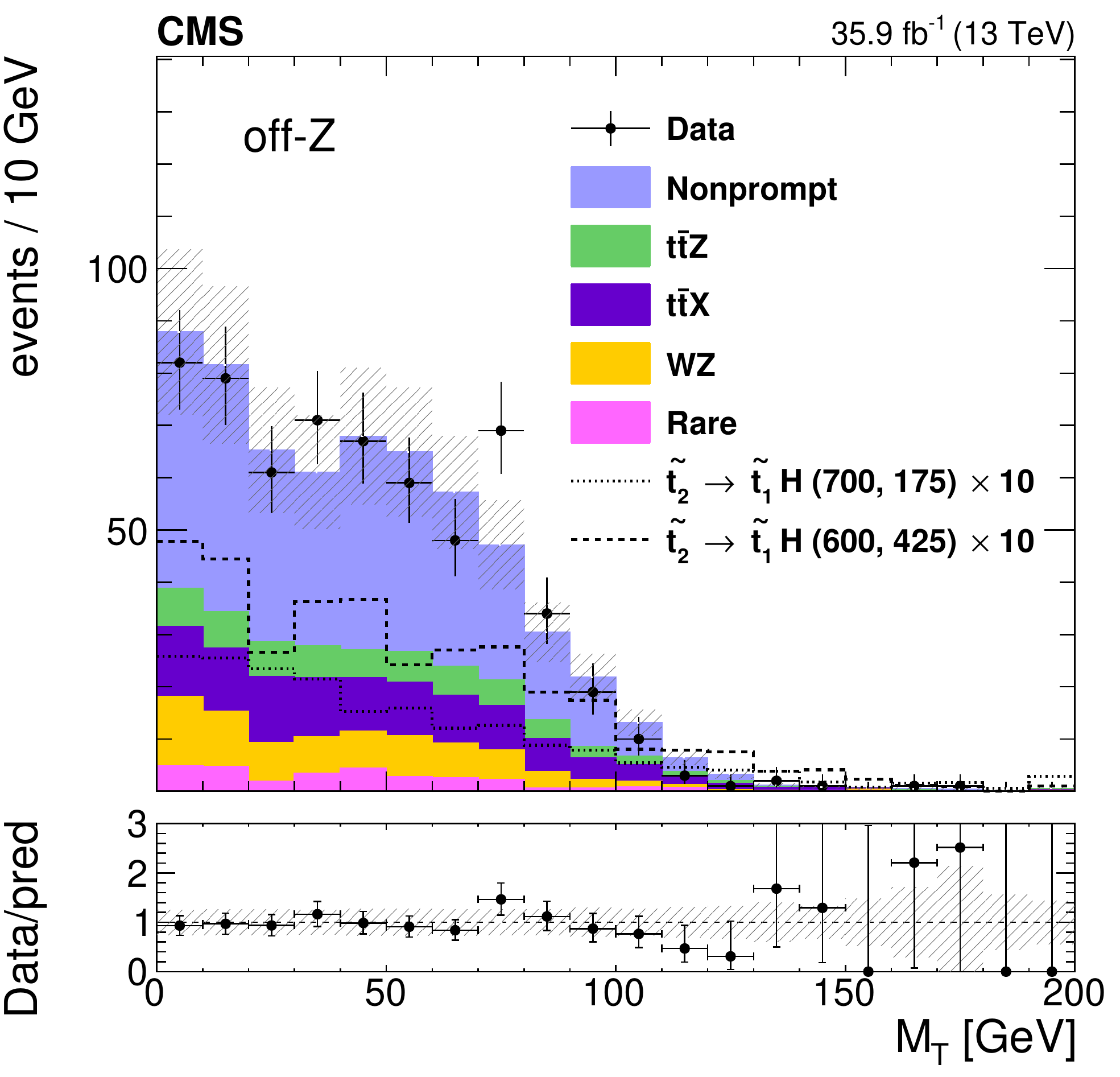}
\includegraphics[width=.32\textwidth]{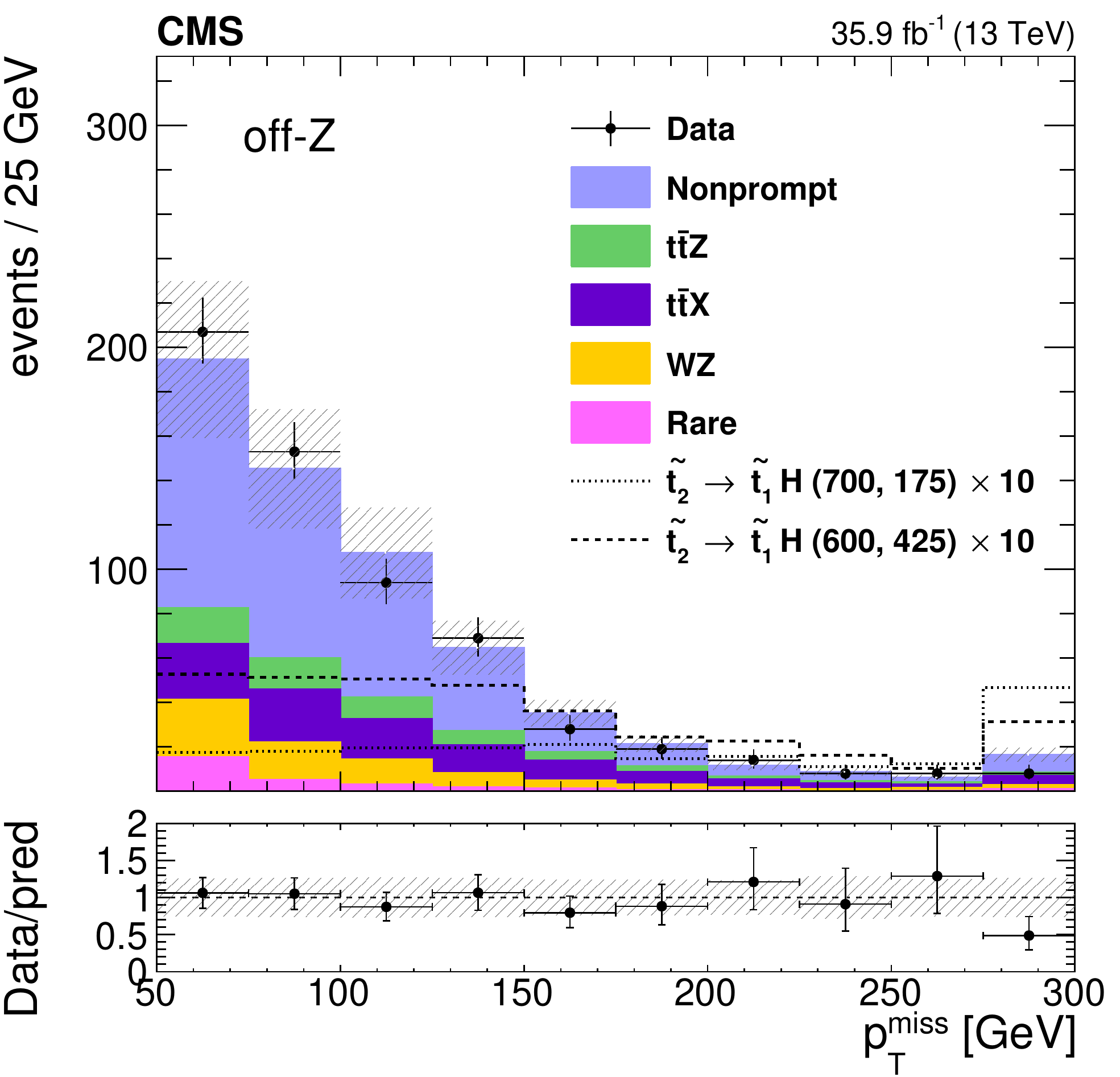}
\includegraphics[width=.32\textwidth]{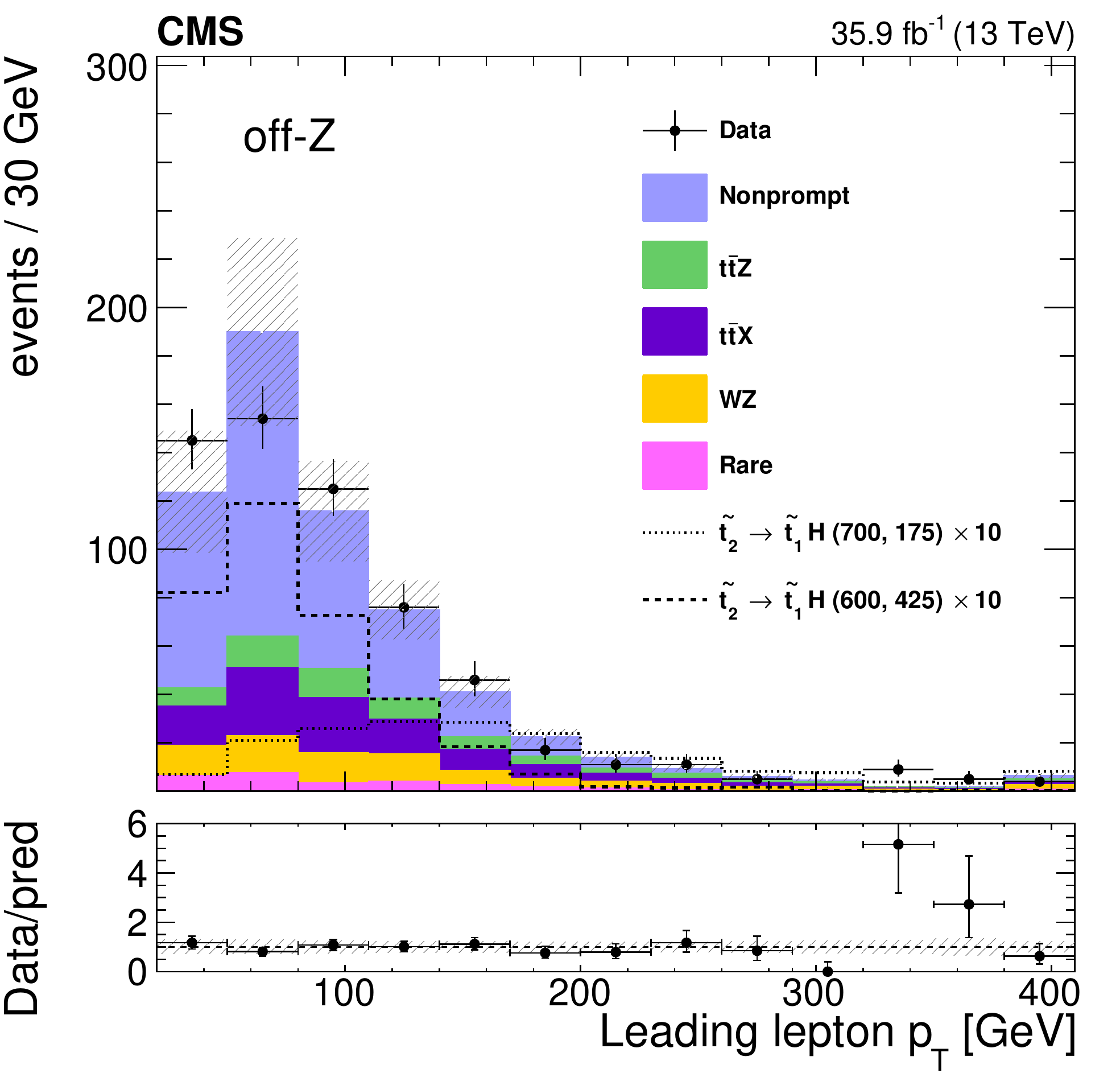}\\
\includegraphics[width=.32\textwidth]{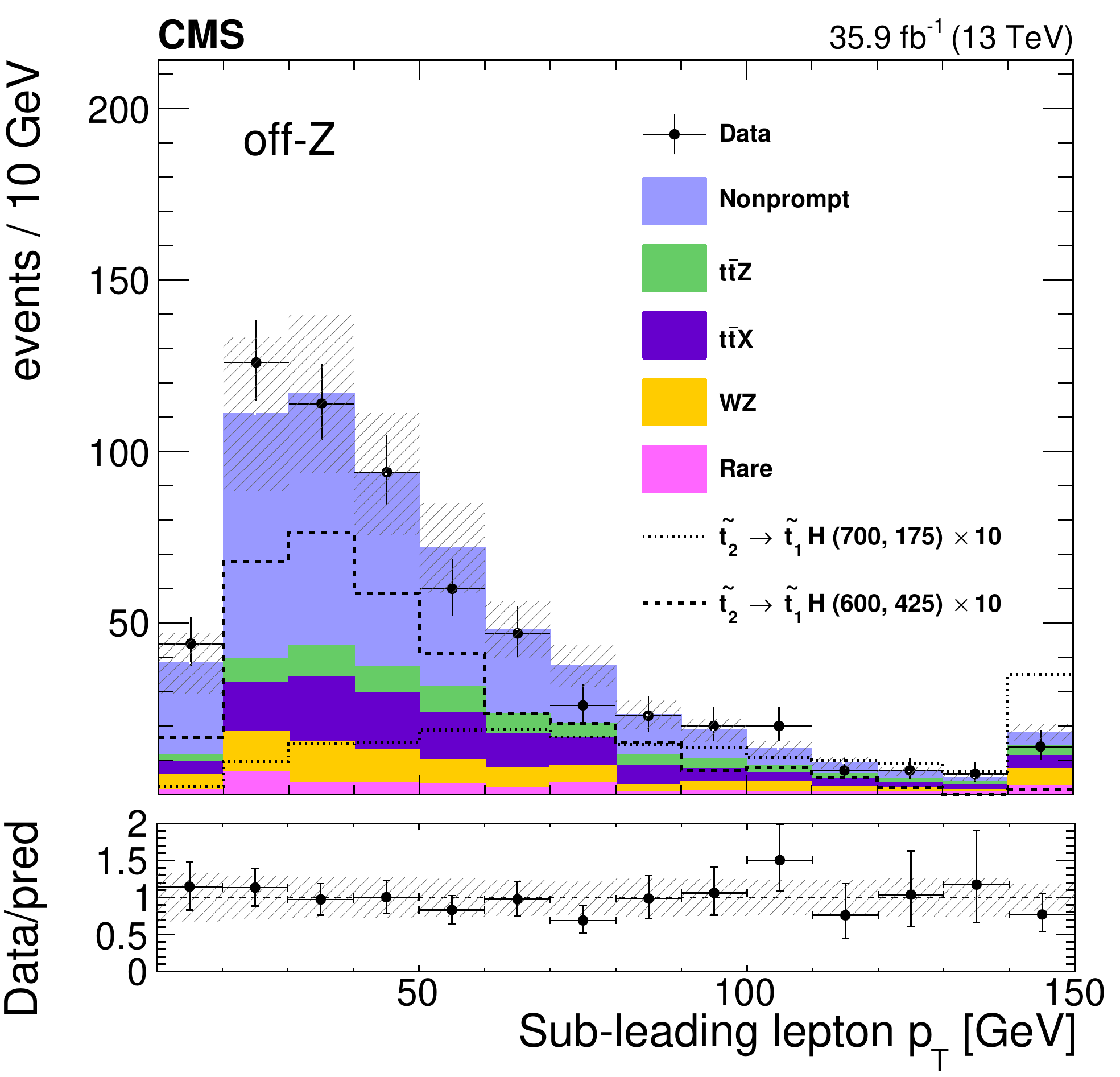}
\includegraphics[width=.32\textwidth]{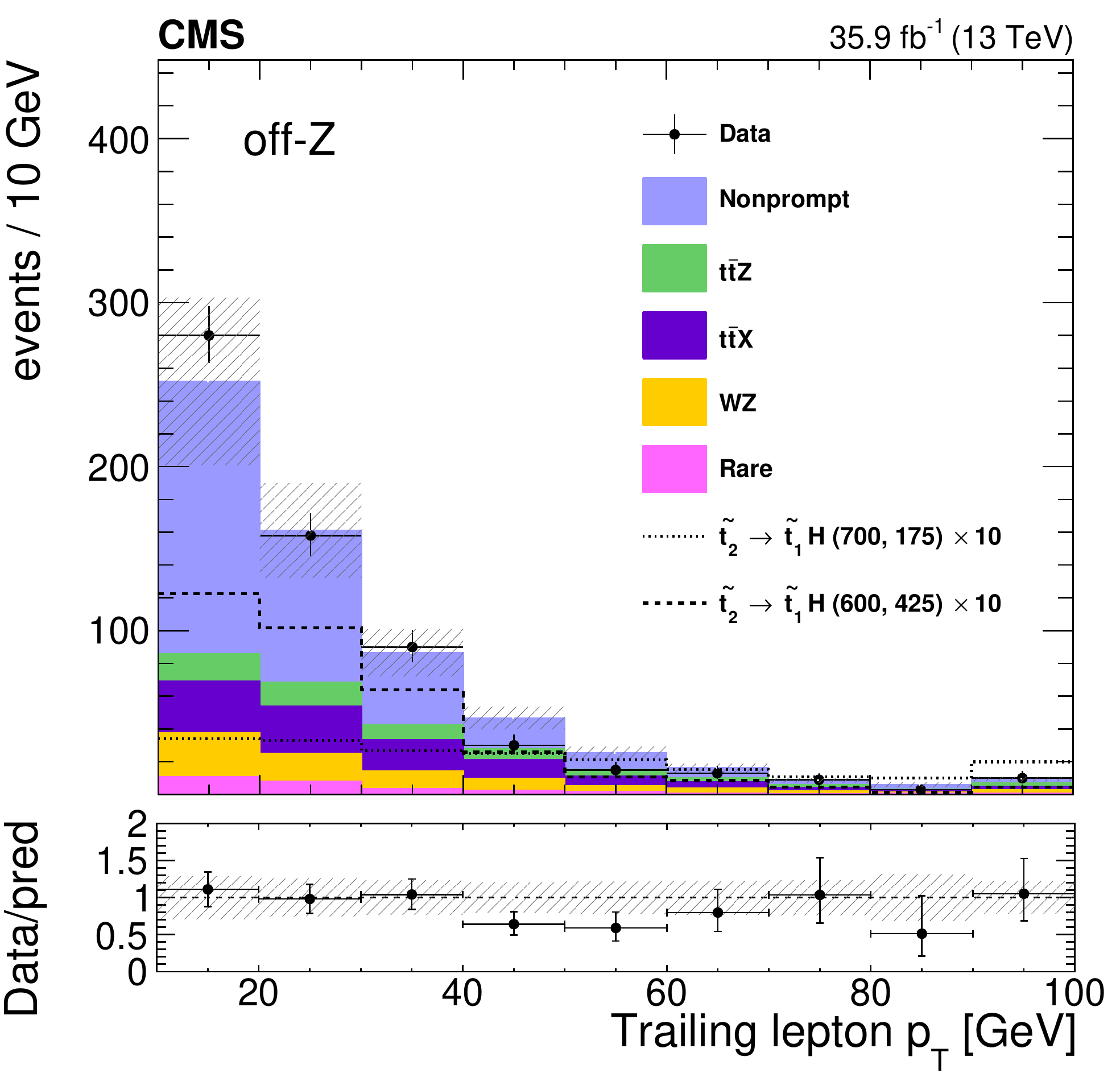}
\includegraphics[width=.32\textwidth]{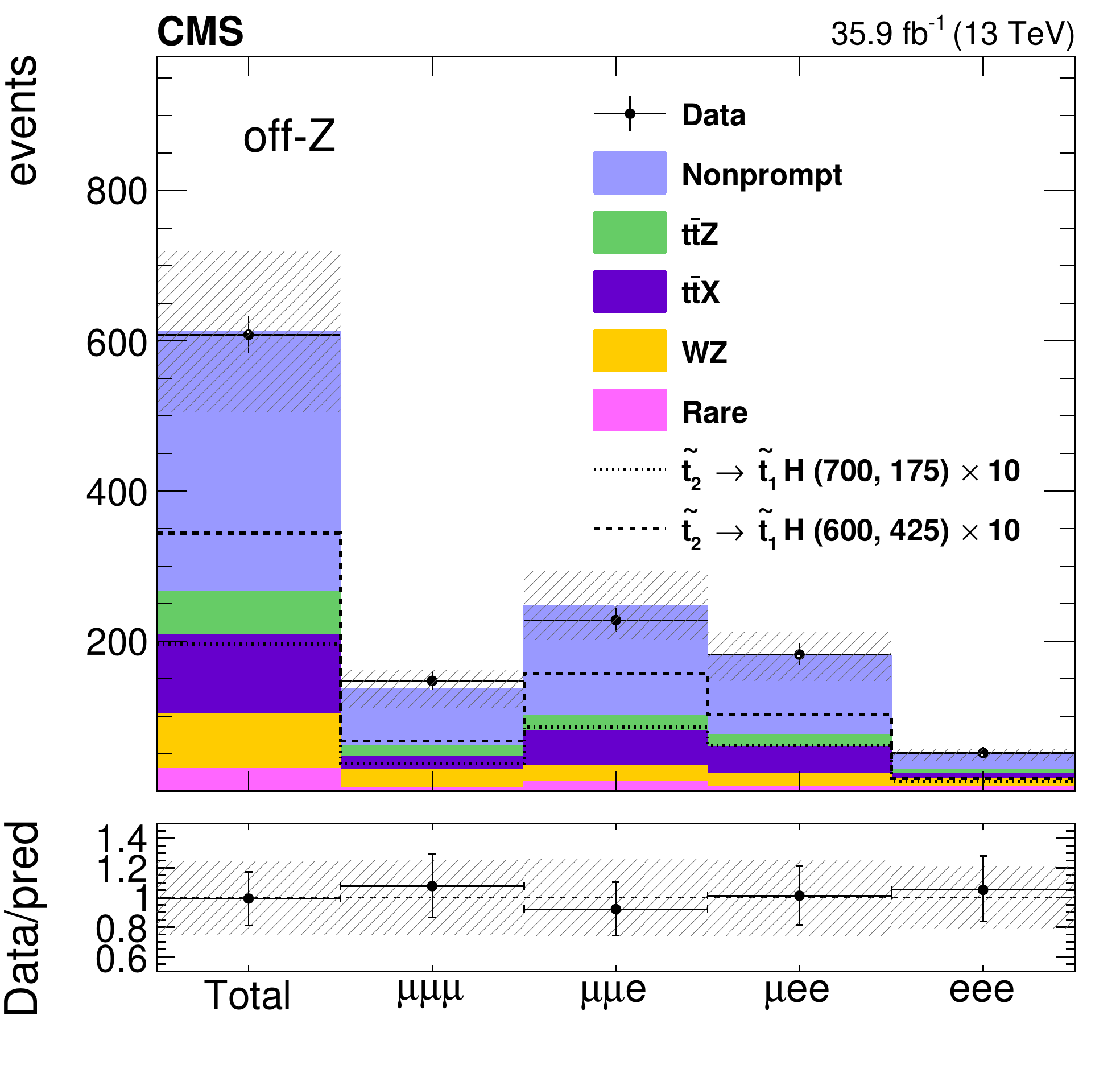}
\caption{Background prediction and the observed event yields in the key
observables for the off-Z baseline selection: the number of
jets and b jets, \HT, \MT, \ptmiss, the lepton \pt spectra and the event yields by flavor category are shown. The background events containing top quark(s) in association with a W, Z or Higgs boson, except \ttZ, or another pair of top quarks are denoted as $\ttbar$X. The last bin includes the overflow events, and the hatched area represents the
statistical and combined systematic uncertainties in the prediction.
The
lower panels show the ratio of the observed and predicted yields
in each bin. For illustration the yields, multiplied by a factor 10, for two signal mass points in the T6ttHZ model, where the $\mathcal{B}(\stoptwo \to \stopone \PH$) = 100\%, are displayed for non-compressed (m(\stoptwo) = 700\GeV and m(\stopone) = 175\GeV) and compressed (m(\stoptwo) = 600\GeV and m(\stopone) = 425\GeV) scenarios. }
\label{fig:resultOffZ}
\end{figure}

\begin{figure}[h]
\centering
\includegraphics[width=.32\textwidth]{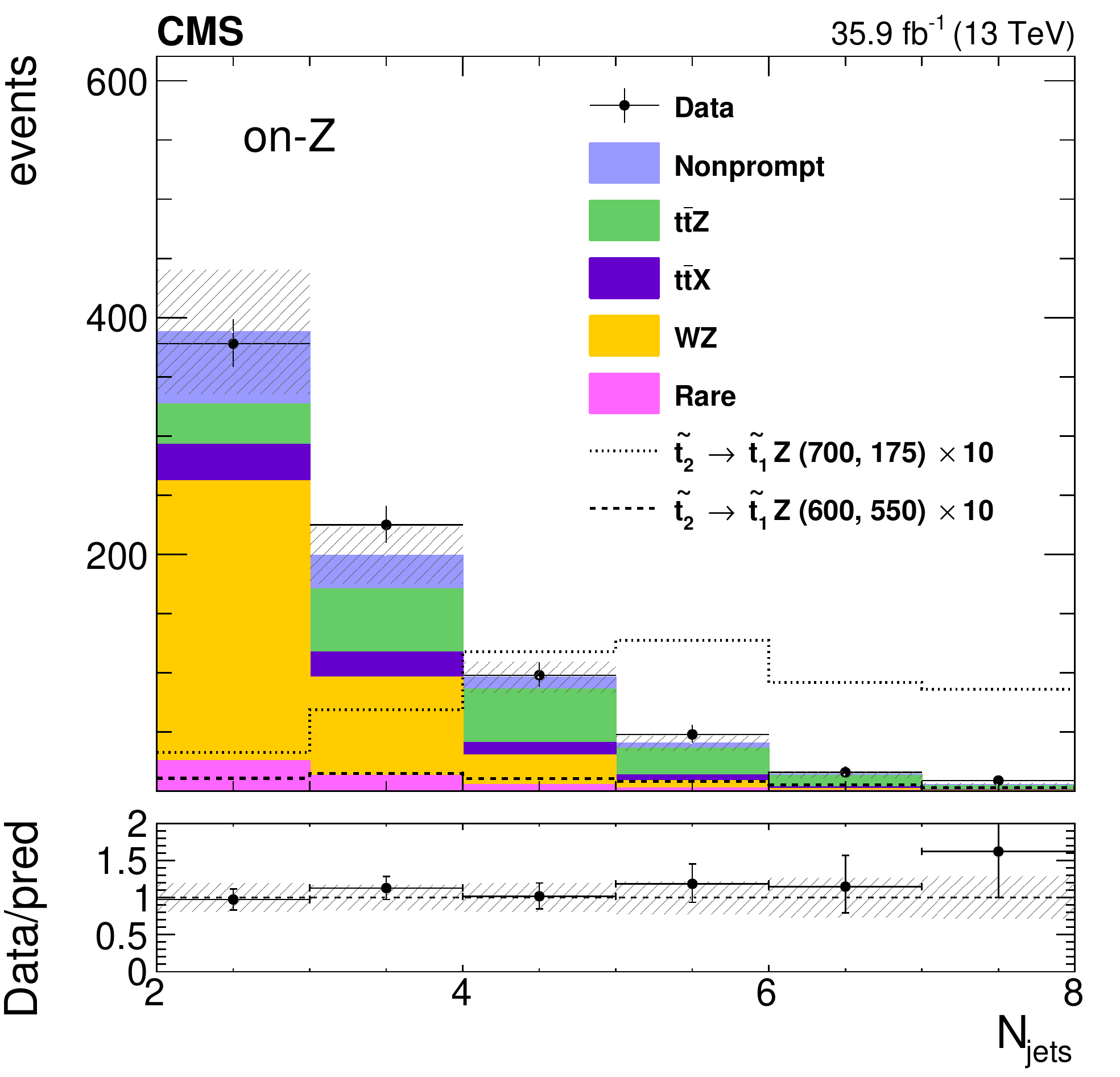}
\includegraphics[width=.32\textwidth]{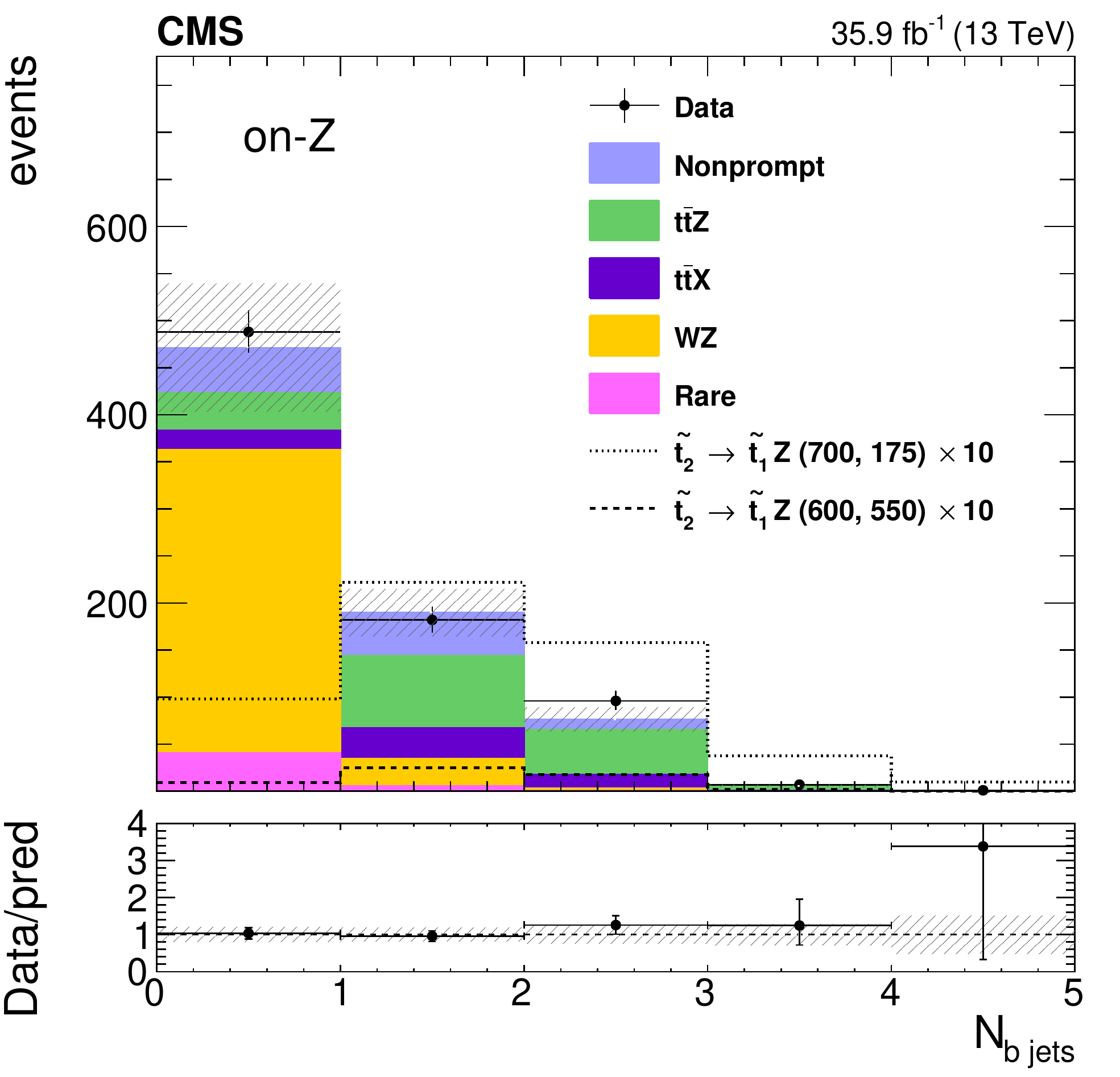}
\includegraphics[width=.32\textwidth]{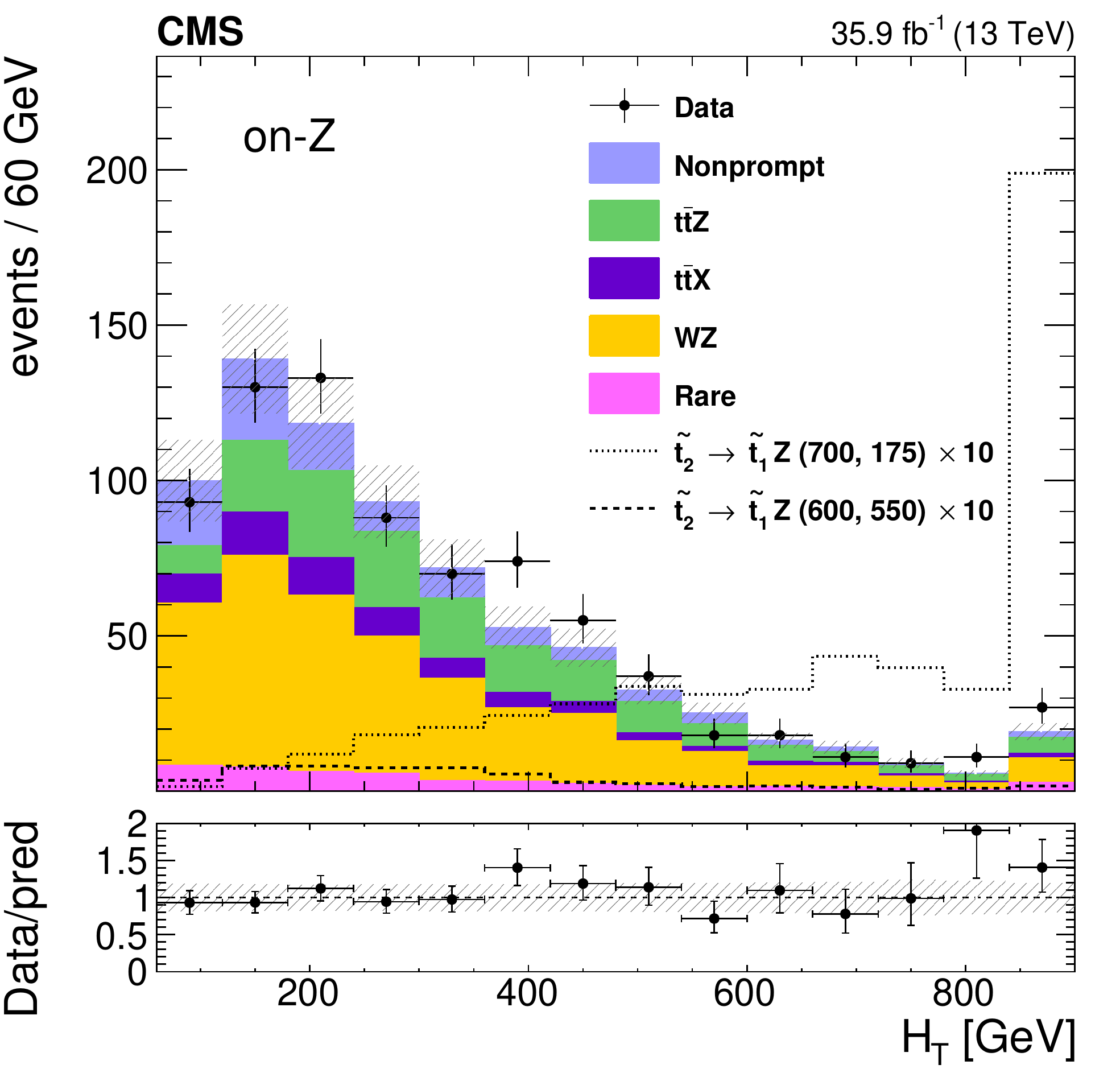}\\
\includegraphics[width=.32\textwidth]{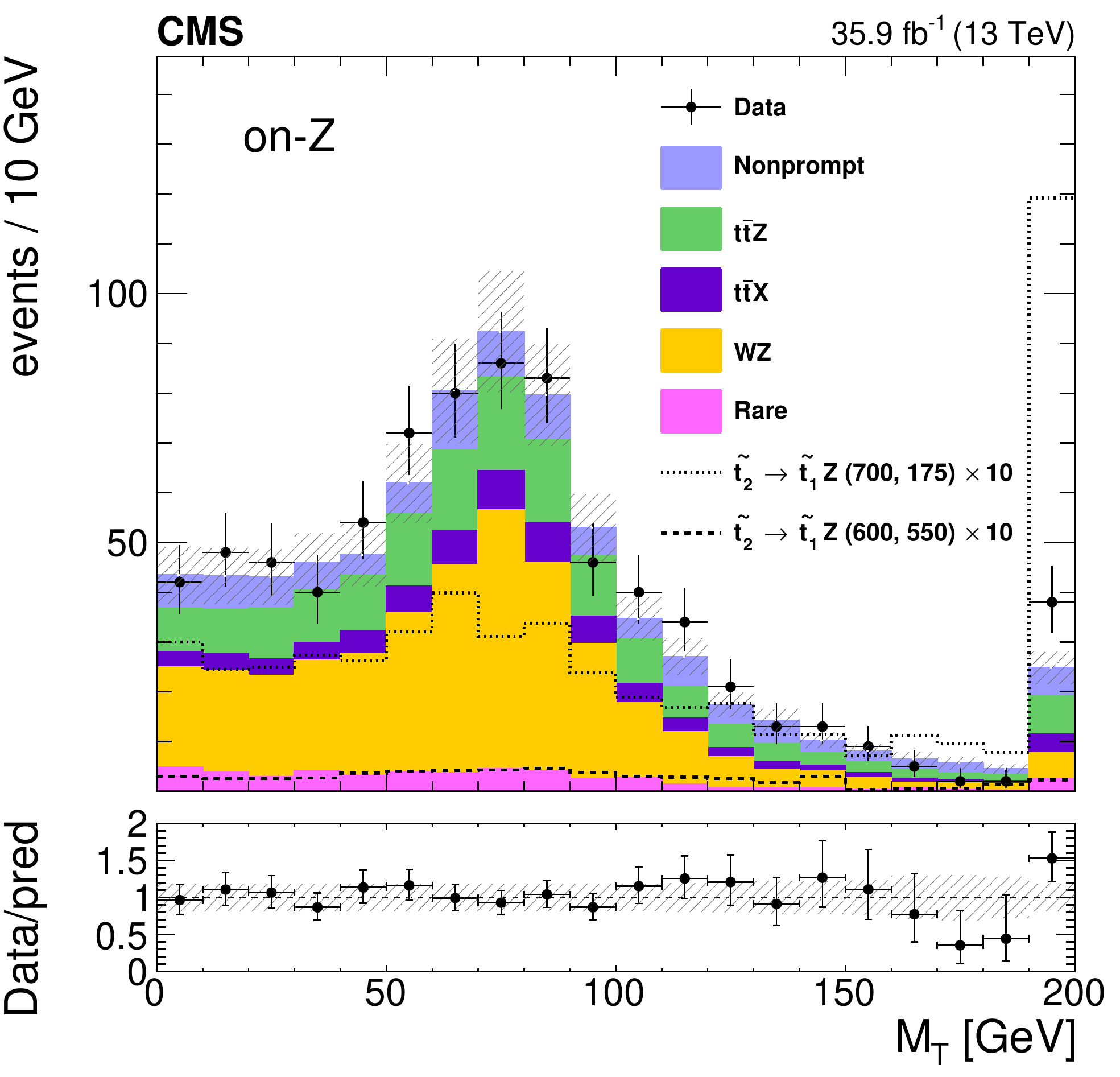}
\includegraphics[width=.32\textwidth]{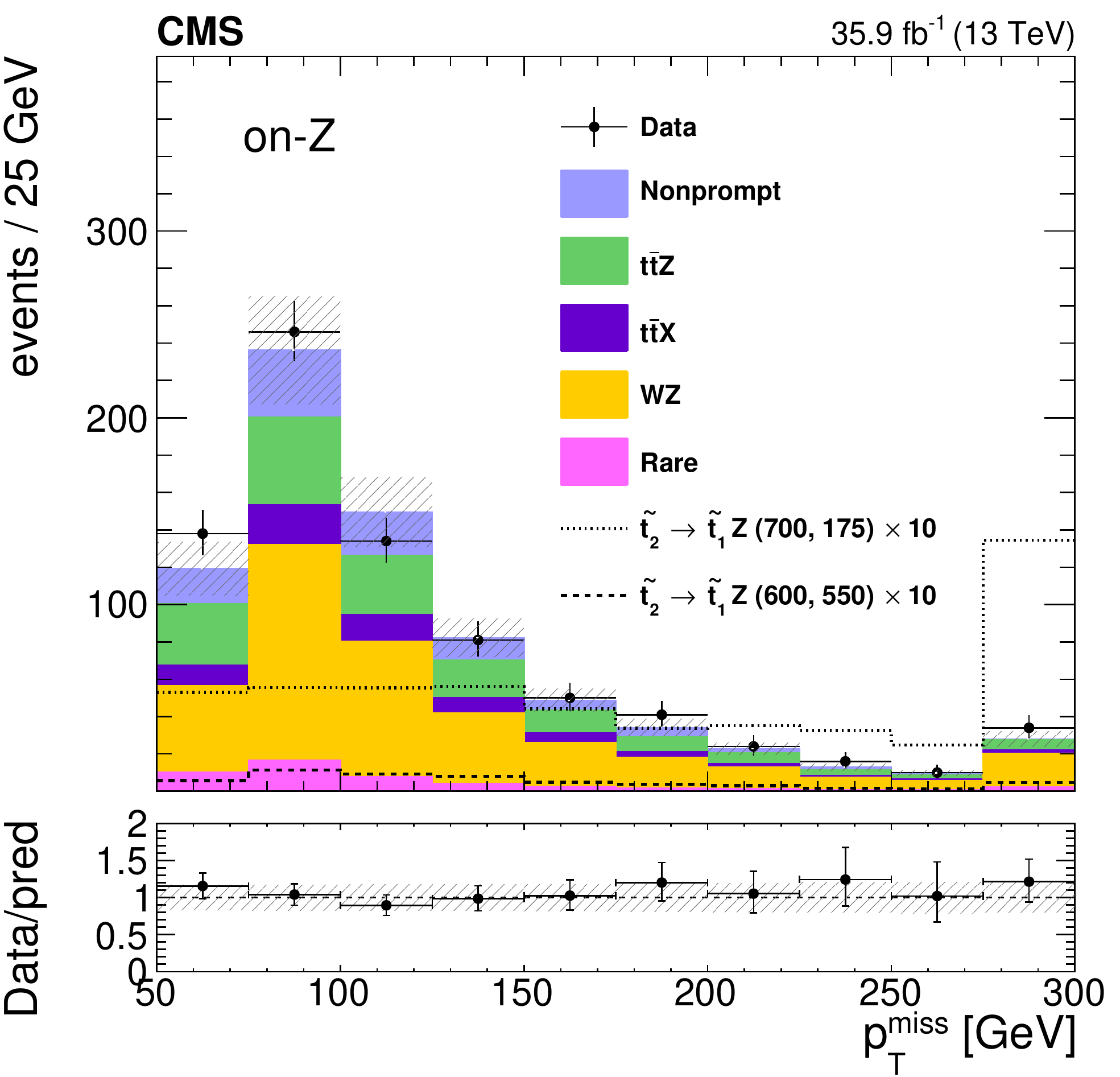}
\includegraphics[width=.32\textwidth]{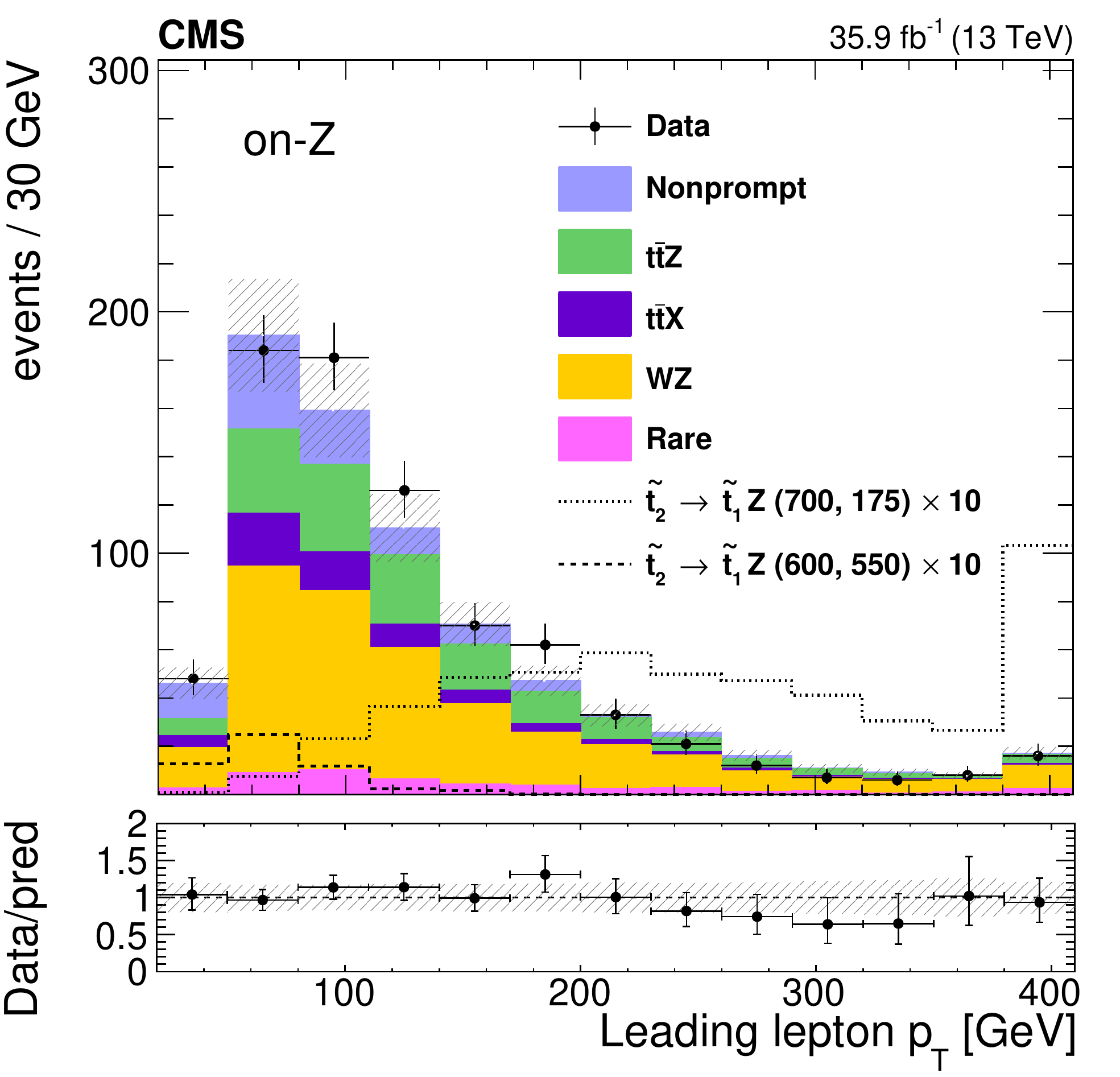}\\
\includegraphics[width=.32\textwidth]{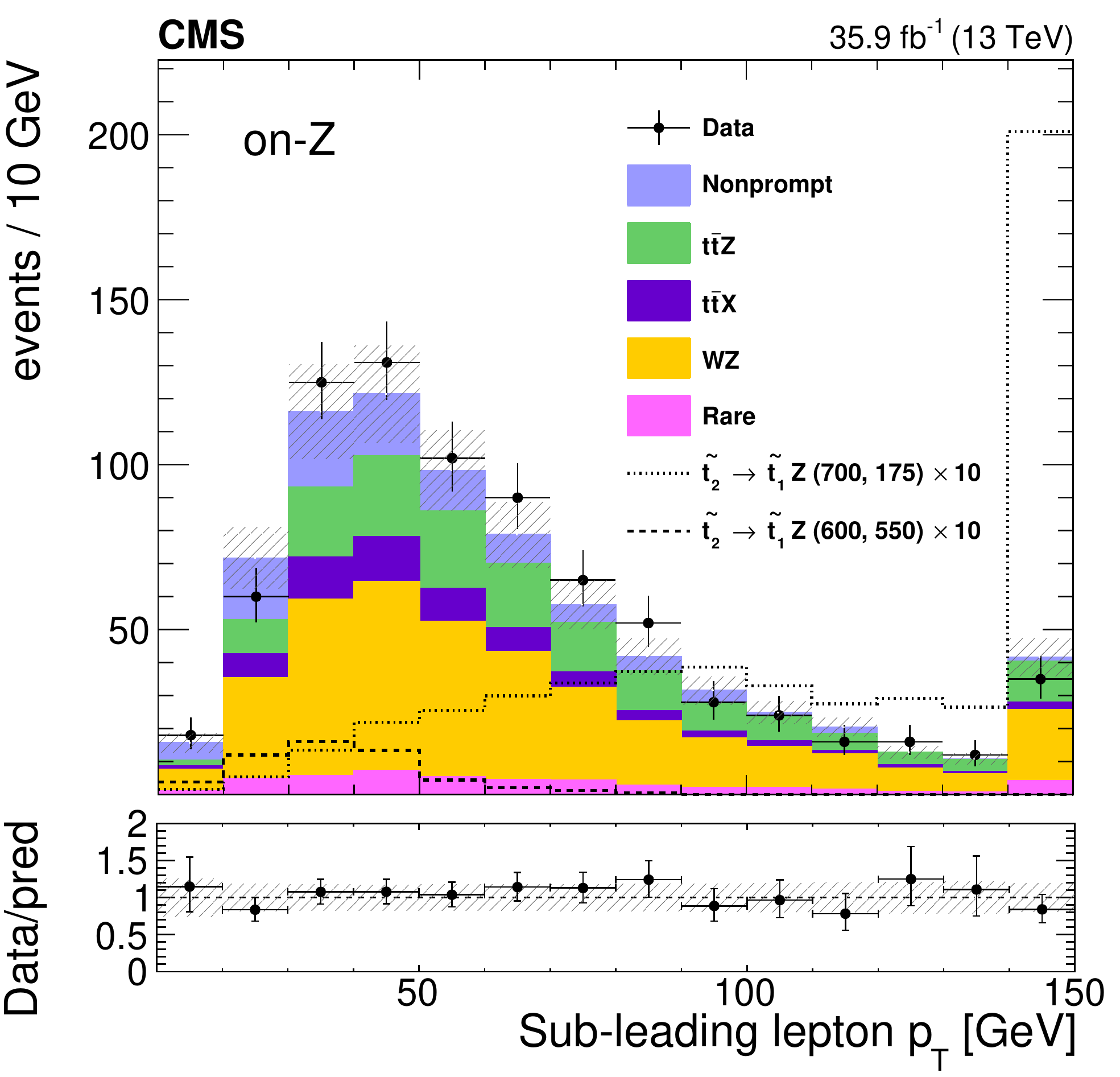}
\includegraphics[width=.32\textwidth]{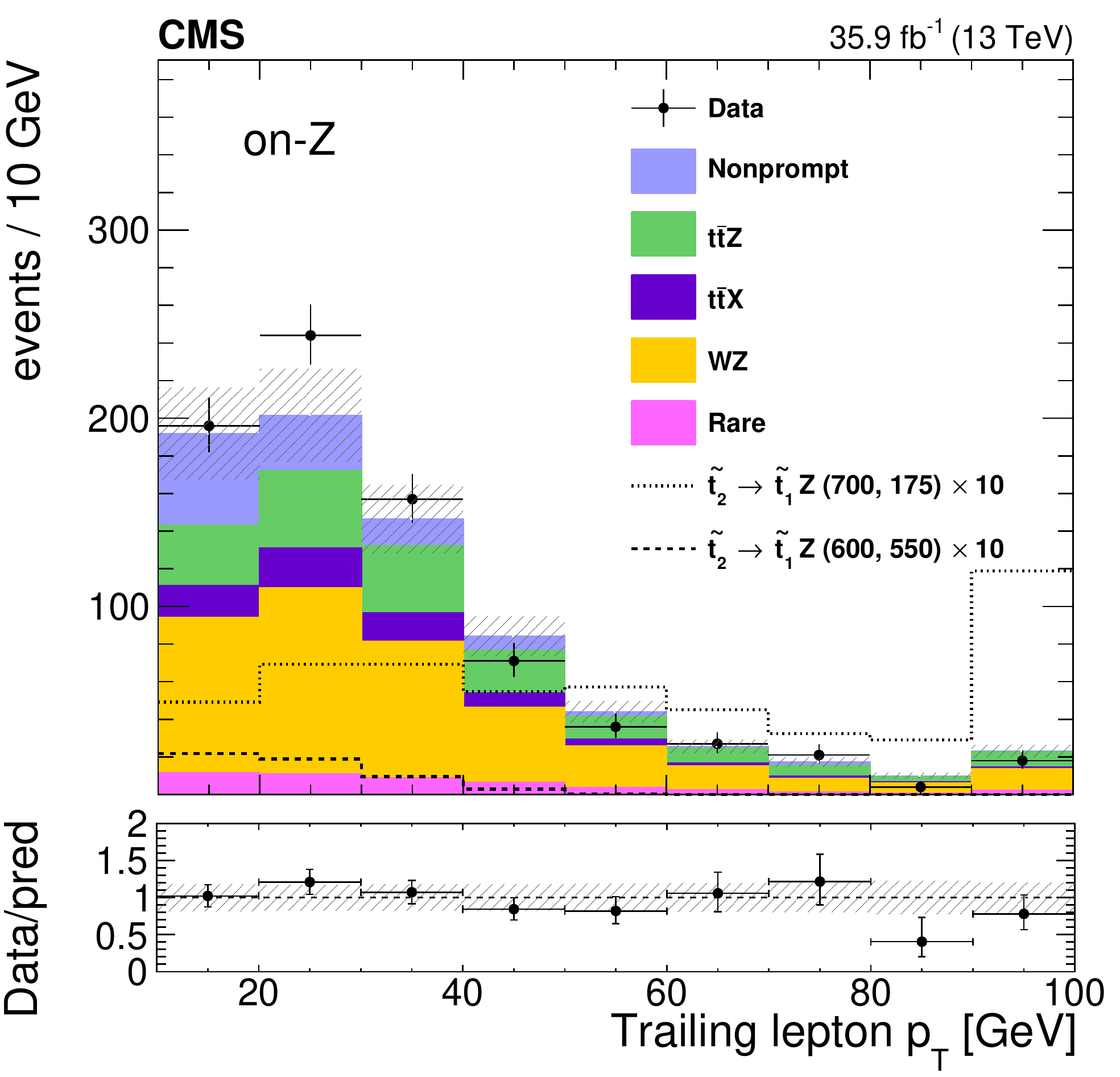}
\includegraphics[width=.32\textwidth]{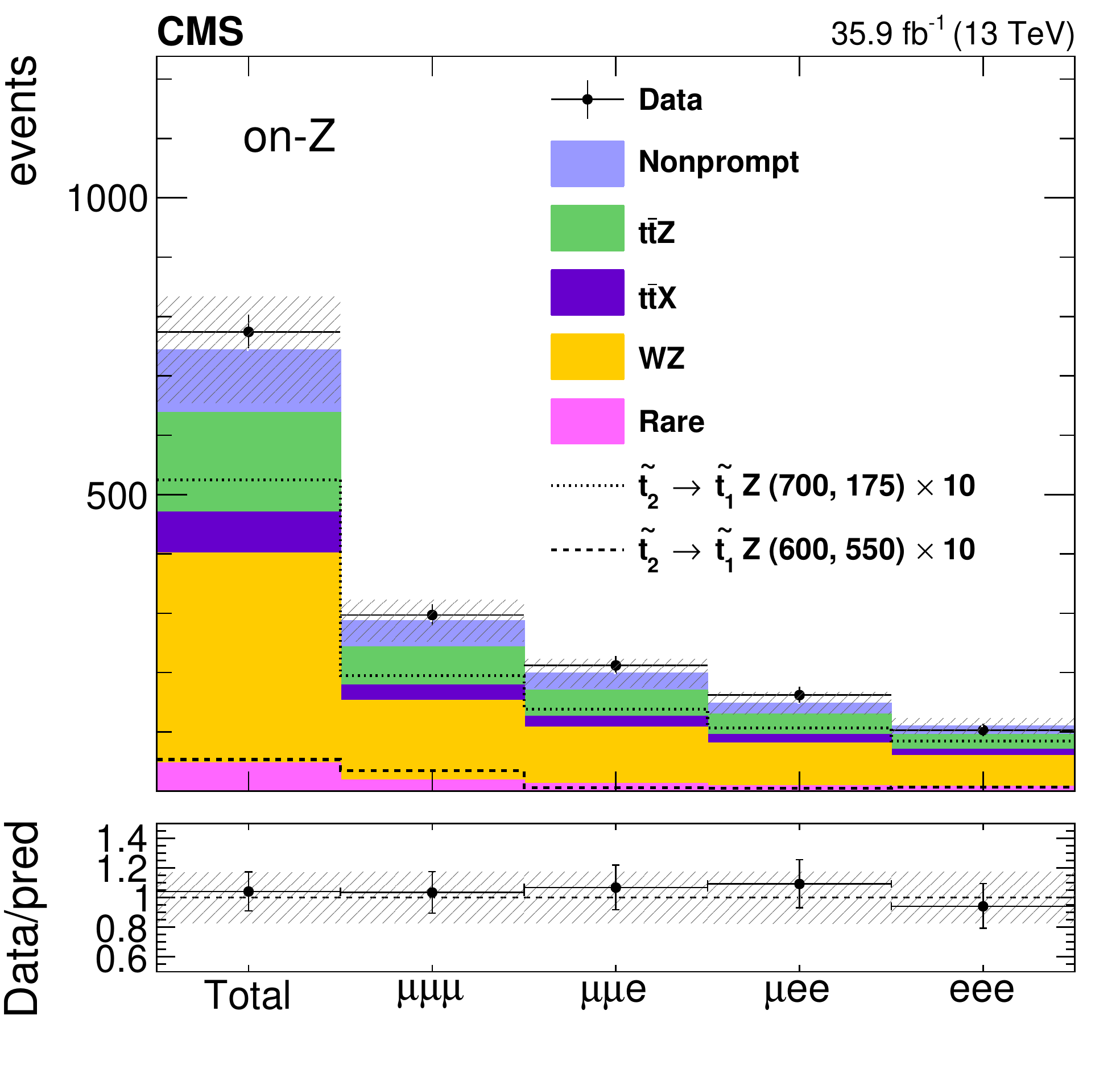}
\caption{Background prediction and the observed event yields in the key
observables of the on-Z baseline selection: the number of
jets and b jets, \HT, \MT, \ptmiss, the lepton \pt spectra and the event yields by flavor category are shown. The background events containing top quark(s) in association with a W, Z or Higgs boson, except \ttZ, or another pair of top quarks are denoted as $\ttbar$X. The last bin includes the overflow events, and the hatched
area represents the combined statistical and systematic uncertainties in
the prediction.
The lower panels show the ratio of the observed
and predicted yields in each bin. For illustration the yields, multiplied by a factor 10, for two signal mass points in the T6ttHZ model, where the $\mathcal{B}(\stoptwo \to \stopone \PZ$) = 100\%, are displayed for non-compressed (m(\stoptwo) = 700\GeV and m(\stopone) = 175\GeV) and compressed (m(\stoptwo) = 600\GeV and m(\stopone) = 550\GeV) scenarios.}
\label{fig:resultOnZ}
\end{figure}

\begin{figure}[h]
\centering
\includegraphics[width=.49\textwidth]{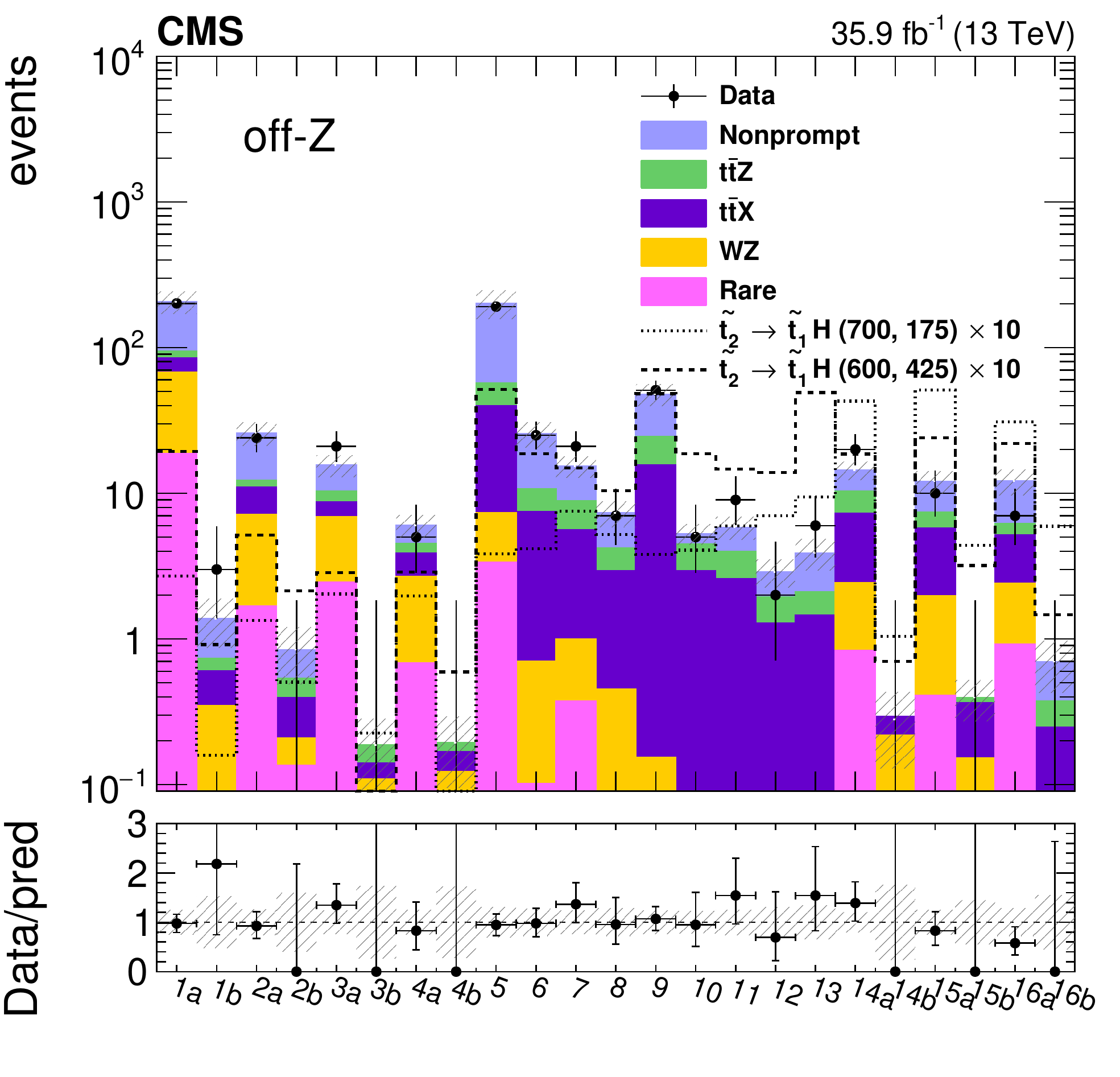}
\includegraphics[width=.49\textwidth]{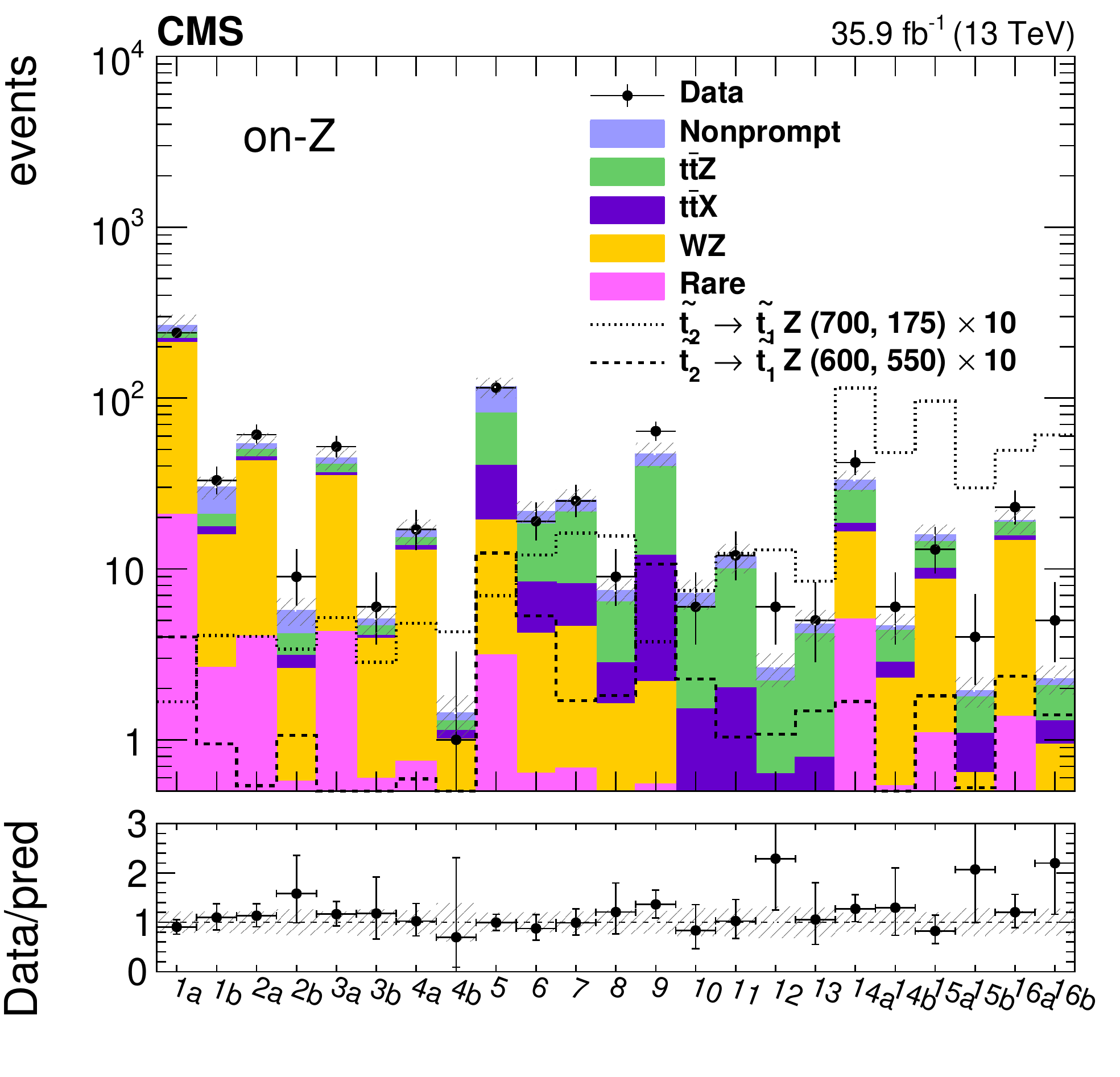}
\caption{Background prediction and observed event yields in the 23 off-Z (left) and the 23 on-Z (right) signal regions. The background events containing top quark(s) in association with a W, Z or Higgs boson, except \ttZ, or another pair of top quarks are denoted as $\ttbar$X. The hatched area represents the statistical and systematic uncertainties on the prediction. The lower panels show the ratio of the observed and predicted yields in each bin. For illustration the yields, multiplied by a factor 10, for \stoptwo $\to$ \stopone H (left) and \stoptwo $\to$ \stopone Z (right) decays are displayed for two signal mass points in the T6ttHZ model to represent compressed and non-compressed scenarios.
}
\label{fig:srs}
\end{figure}

Comparisons between data and the predicted background of the distributions of the four event observables used
for signal region categorization, namely $\HT$, $\ptmiss$, $\MT$ and
$\Nbjets$, as well as the lepton $\pt$ spectra, the lepton flavor
composition, and the event jet multiplicity
are shown in
Fig.~\ref{fig:resultOffZ} (Fig.~\ref{fig:resultOnZ}) for
events satisfying the selection criteria of the off-Z (on-Z).
Figure~\ref{fig:srs} graphically presents a summary of the
predicted background and observed event yields in the individual
SR bins. The same information is also presented in
Tables~\ref{tab:offZfull} and~\ref{tab:onZfull} for the off-Z and on-Z
regions, respectively. Table \ref{tab:SSRfull} represents the yields
in the SSRs.

The number of events observed in data is found to be consistent with the predicted background yields in all 46 SRs.
The results of the search are interpreted by setting limits on superpartner masses using simplified models.
For each mass point, the observations, background predictions, and expected signal yields from all on-Z and off-Z search regions are combined to extract the minimum cross section that can be excluded at a 95\% CL using the CL$_\text{s}$ method~\cite{Junk:1999kv,Read:2002hq,ATLAS:1379837}, in which asymptotic approximations for the distribution of the test-statistic, which is a ratio of profiled likelihoods, are used~\cite{Cowan:2010js}. Log-normal nuisance parameters are used to describe the uncertainties listed in Section~\ref{sec:systematics}.

\begin{figure}[!hbtp]
\centering
\includegraphics[width=.49\textwidth]{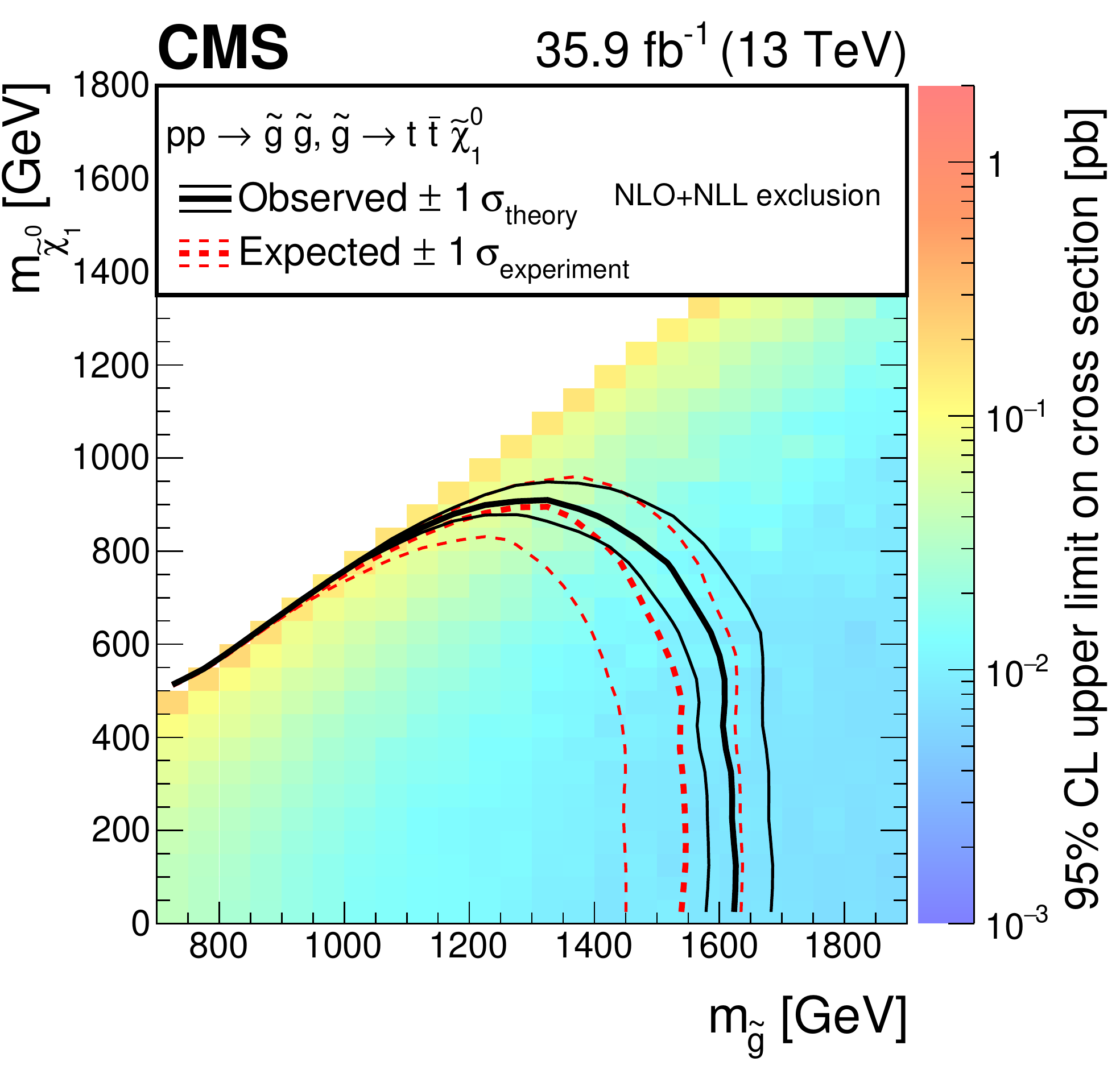}
\includegraphics[width=.49\textwidth]{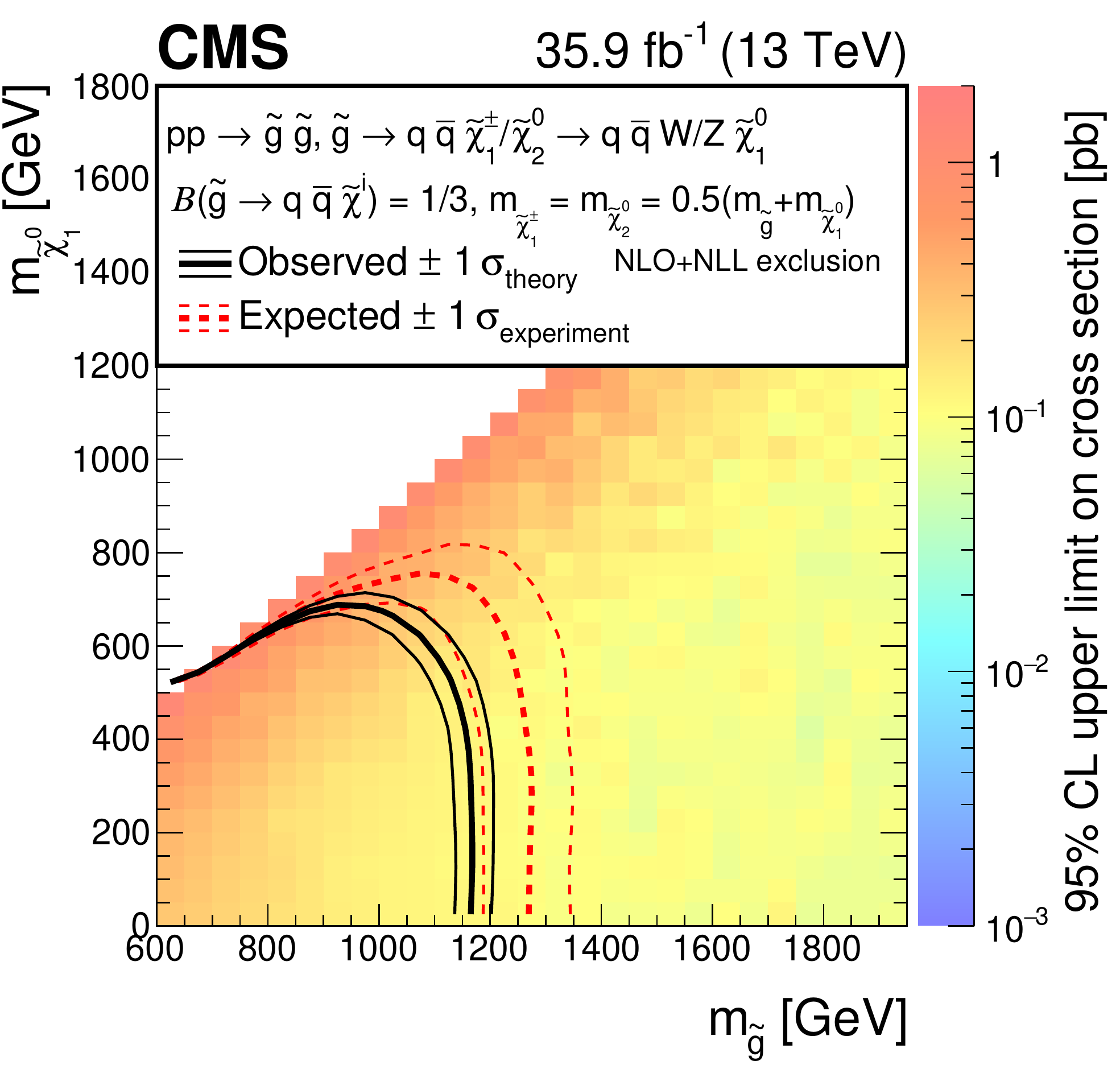}
\caption{Cross section upper limits at 95\% CL in the $m_{\chiz_1}$ versus $m_{\sGlu}$ plane for T1tttt (left) and T5qqqqVV (right) simplified models. For the latter model the branching fraction of gluino decay to neutralino or chargino is equal to 1/3 and $m_{\chiplmin_{1}} = m_{\chiz_2} = 0.5(m_{\gluino} + m_{\lsp})$. The excluded regions are to the left and below the observed and expected limit curves. The color scale indicates the excluded cross section at a given point in the mass plane. }
\label{fig:excl1}
\end{figure}

The limits are shown in Fig.~\ref{fig:excl1} for the T1tttt model (left) and for the T5qqqqVV model (right).
In the T5qqqqVV model each gluino decays to a pair of light quarks and a neutralino ($\PSGcz_2$) or chargino ($\PSGcpm_1$),
followed by the decay of that neutralino or chargino to a W or Z boson, respectively, and an LSP (Fig.~\ref{fig:diagrams}, top right).
The probability for the decay to proceed via the $\PSGcp_1$, $\PSGcm_1$, or $\PSGcz_2$ is taken to be 1/3 for each case.
In this scenario, the second neutralino \chitz and chargino are assumed to be mass-degenerate, with masses equal to $0.5(m_{\sGlu} + m_{\chiz_1})$.

\begin{figure}[!hbtp]
\centering
\includegraphics[width=.49\textwidth]{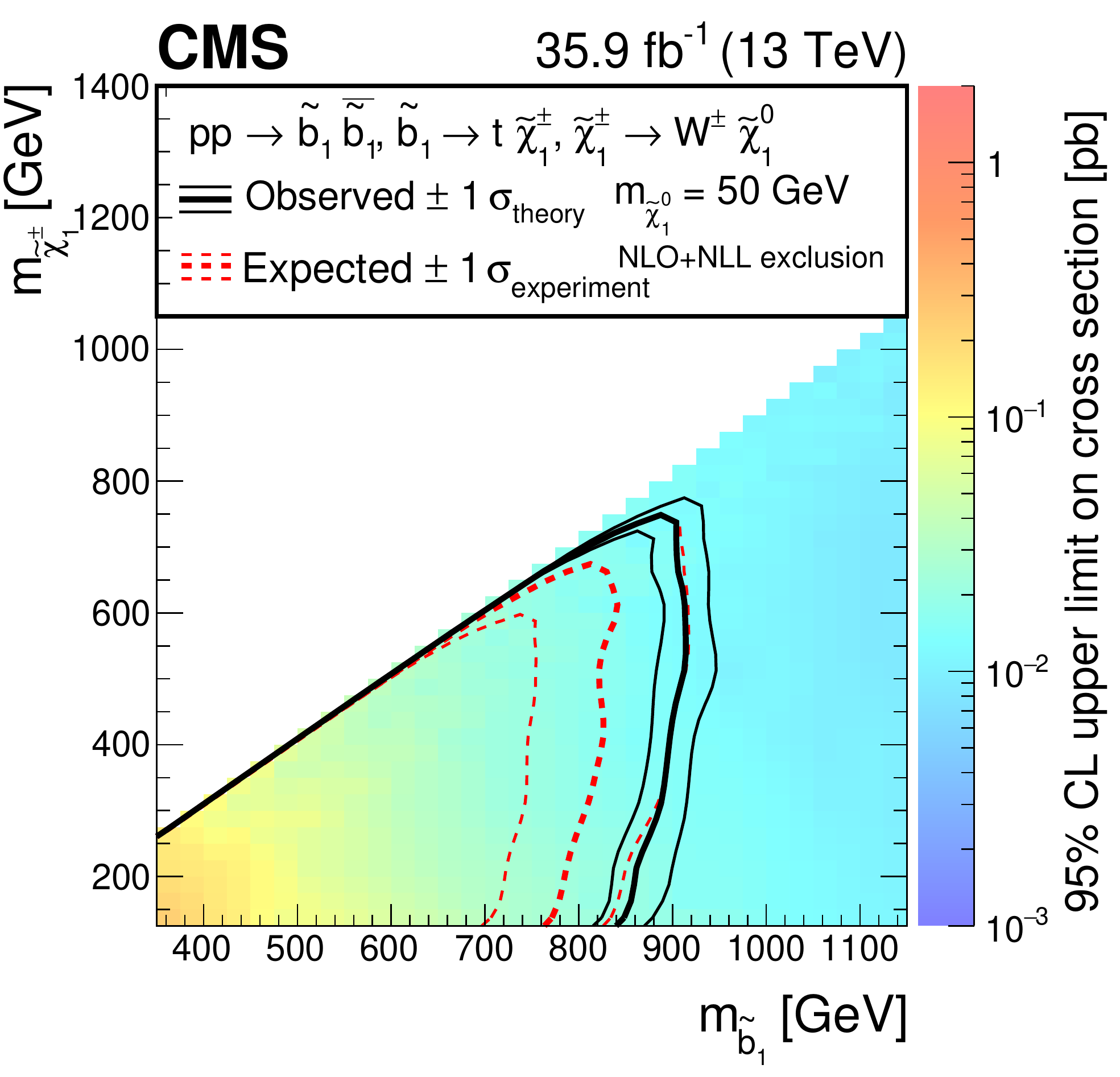}
\caption{Cross section upper limits at 95\% CL in the $m_{\chiplmin_1}$ versus $m_{\sbottom_{1}}$ plane for T6ttWW simplified model. The mass of the neutralino is set to 50\GeV.
The descriptions of the excluded regions and color scale are the same as in Fig.~\ref{fig:excl1}.
}
\label{fig:excl2}
\end{figure}

\begin{figure}[!hbtp]
\centering
\includegraphics[width=.49\textwidth]{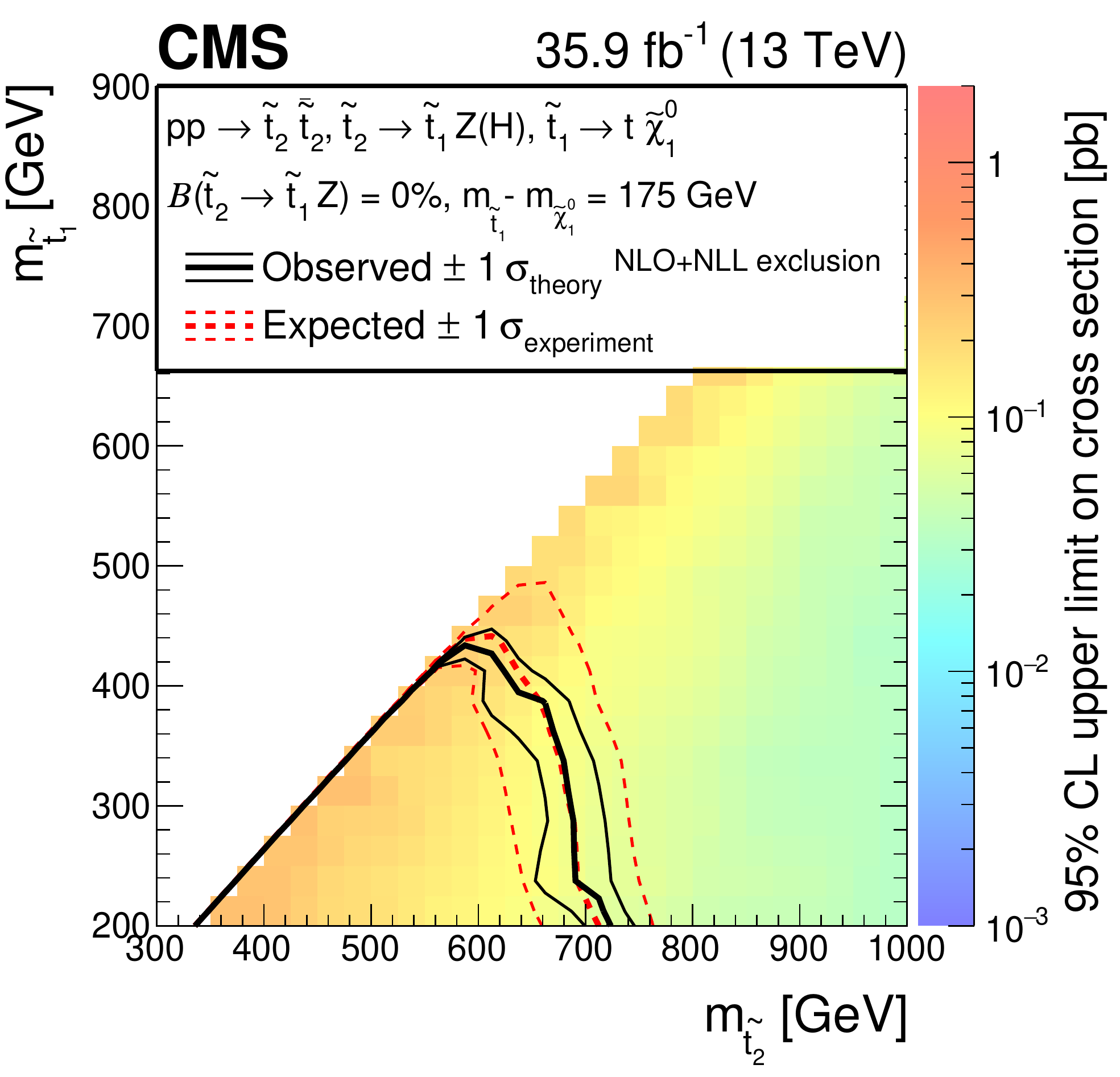}
\includegraphics[width=.49\textwidth]{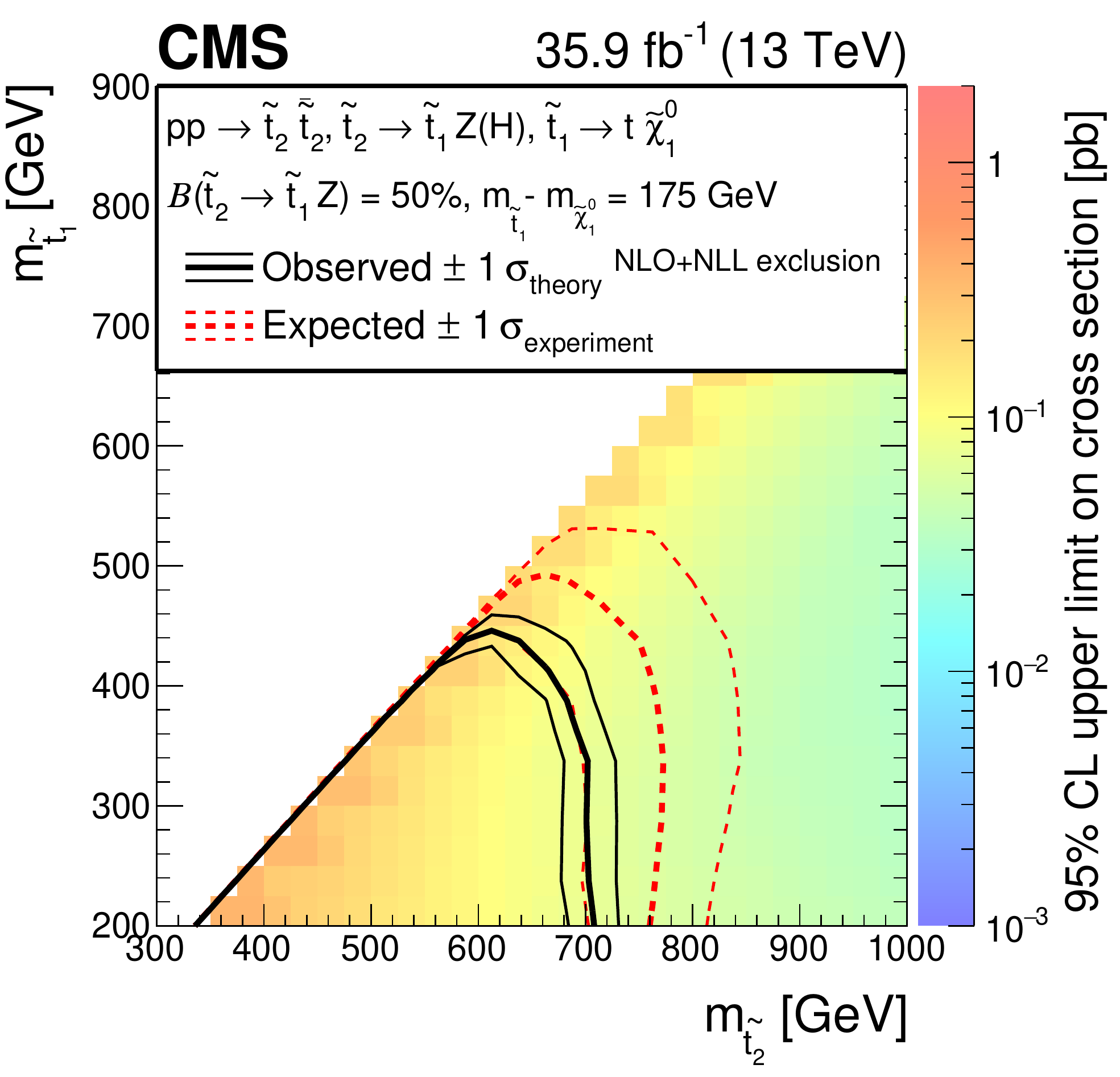}
\includegraphics[width=.49\textwidth]{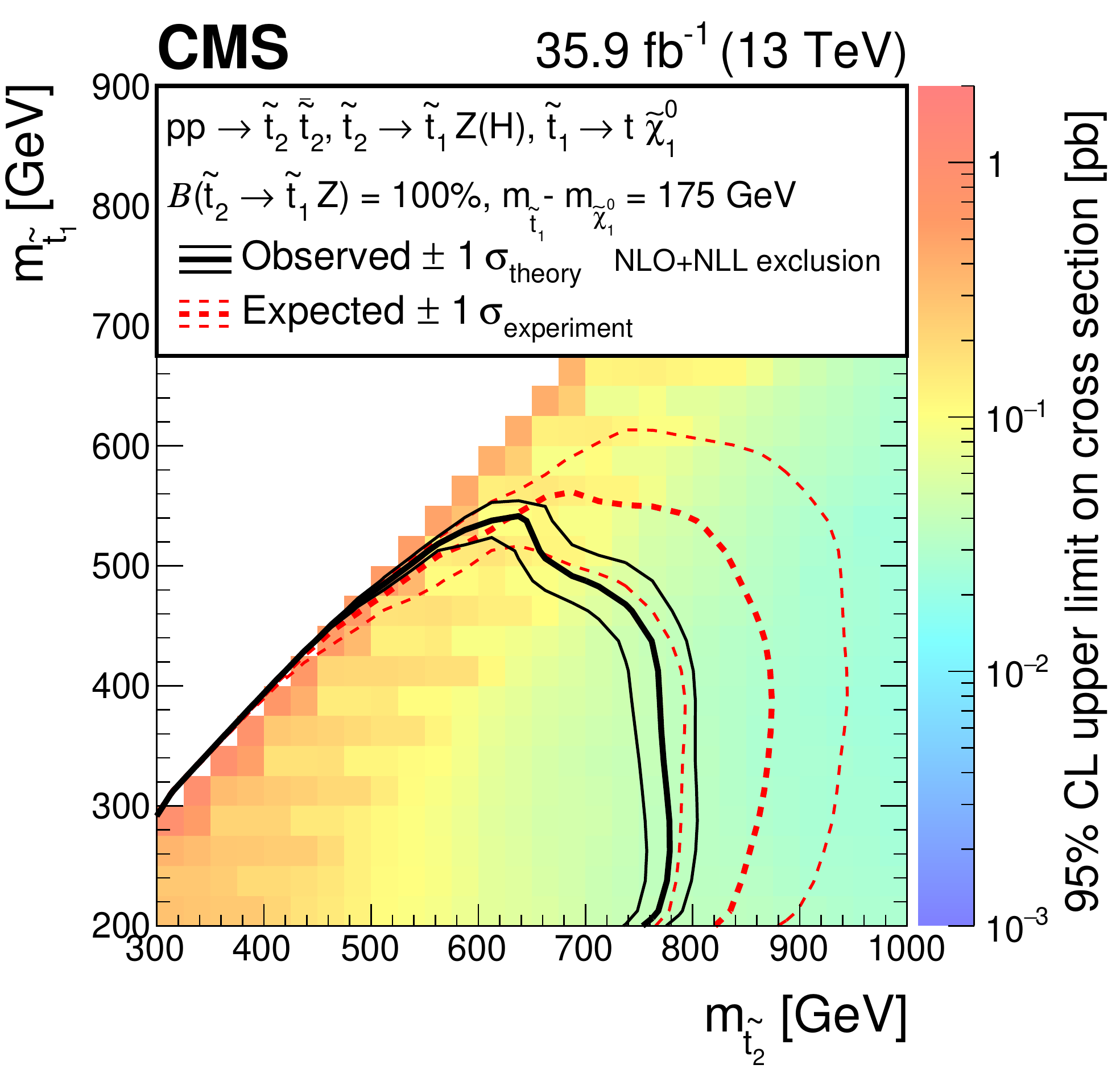}
\caption{Cross section upper limits at 95\% CL in the $m_{\stopone}$ versus $m_{\stoptwo}$ plane for T6ttHZ simplified model. Different branching fractions of the decay $\stoptwo \to\stopone \PZ$ are considered: 0\% (top left), 50\% (top right), and 100\% (bottom). The mass difference between the lighter top squark (\stopone) and a neutralino is close to the mass of the top quark. The descriptions of the excluded regions and color scale are the same as in Fig.~\ref{fig:excl1}.
}
\label{fig:excl3}
\end{figure}

The limits on the bottom squark pair production cross section are shown in Fig.~\ref{fig:excl2}. In this model,
the mass of the LSP is set to 50\GeV. Finally, the limits on the $\stoptwo$ pair production cross section
are shown in Fig.~\ref{fig:excl3}. In this scenario, the mass difference between the $\stopone$ and the LSP
is set to 175\GeV, the $\stopone$ decays via a top quark to LSP, and the $\stoptwo$ decays via a $\PZ$ or Higgs boson to $\stopone$. We consider the reference values $\mathcal{B}(\PSQtDt\to\PSQtDo\PZ) = 0$,
50, and 100\%; the sensitivity is diminished for the $\PSQtDo\PH$ final
state because of the additional branching factors for Higgs cascade
decays to electrons or muons via gauge bosons or tau leptons.

\begin{table}[b!]
\topcaption{Expected and observed yields in the off-Z search regions. The first uncertainty states the statistical uncertainty, while the second represents the systematic uncertainty.}
\label{tab:offZfull}
\begin{center}
\resizebox{0.9\linewidth}{!}{
\renewcommand{\arraystretch}{1.3}
\begin{tabular}{ccccccc}\hline
$\Nbjets$                           & $\HT$ [\GeVns{}]                        & $\ptmiss$ [\GeVns{}]                & $\MT$ [\GeVns{}]                     &  Expected [events]                                                     &  Observed [events]                     & SR          \\ \hline
\multirow{8}{*}{0}        &  \multirow{4}{*}{\x60--400}          & \multirow{2}{*}{\x50--150}                 & ${<}120$                          &  $206\pm6\pm35$ & $201$ & SR1a  \\ \cline{4-7}
&                                    &                                          & ${\geq}1 20$                      &  $1.4\pm0.5\pm0.2$ & $3$ & SR1b  \\ \cline{3-7}
&                                    & \multirow{2}{*}{150--300}                & ${<} 120$                          &  $25.9\pm2.1\pm4.3$ & $24$ & SR2a  \\ \cline{4-7}
&                                    &                                          & ${\geq}1 20$                      &  $0.84\pm0.34\pm0.12$ & $0$ & SR2b  \\ \cline{2-7}
&  \multirow{4}{*}{400-600}         & \multirow{2}{*}{\x50--150}                 & ${<} 120$                          &  $15.6\pm1.6\pm2.1$ & $21$ & SR3a  \\ \cline{4-7}
&                                    &                                          & ${\geq}1 20$                      &  $0.19\pm0.09\pm0.02$ & $0$ & SR3b  \\ \cline{3-7}
&                                    & \multirow{2}{*}{150--300}                & ${<} 120$                          &  $6.0\pm0.8\pm0.7$ & $5$ & SR4a  \\ \cline{4-7}
&                                    &                                          & ${\geq}1 20$                      &  $0.19\pm0.09\pm0.04$ & $0$ & SR4b  \\ \hline
\multirow{4}{*}{1}        &  \multirow{2}{*}{\x60--400}          & \x50--150                                   & \multirow{4}{*}{Inclusive}      &  $202\pm6\pm44$ & $191$ & SR5  \\ \cline{3-3}\cline{5-7}
&                                    & 150--300                                  &                                  &  $25.6\pm1.9\pm4.6$ & $25$ & SR6  \\ \cline{2-3}\cline{5-7}
&  \multirow{2}{*}{400-600}         & \x50--150                                   &                                  &  $15.4\pm1.3\pm2.2$ & $21$ & SR7  \\ \cline{3-3}\cline{5-7}
&                                    & 150--300                                  &                                  &  $7.3\pm1\pm1.1$ & $7$ & SR8  \\ \hline
\multirow{4}{*}{2}        &  \multirow{2}{*}{\x60--400}          & \x50--150                                   & \multirow{4}{*}{Inclusive}      &  $47.7\pm2.8\pm7.6$ & $51$ & SR9  \\ \cline{3-3}\cline{5-7}
&                                    & 150--300                                  &                                  &  $5.3\pm0.5\pm0.6$ & $5$ & SR10  \\ \cline{2-3}\cline{5-7}
&  \multirow{2}{*}{400-600}         & \x50--150                                   &                                  &  $5.8\pm0.7\pm0.8$ & $9$ & SR11  \\ \cline{3-3}\cline{5-7}
&                                    & 150--300                                  &                                  &  $2.9\pm0.5\pm0.4$ & $2$ & SR12  \\ \hline
$ {\geq}3$                 &  \x60--600                            & \x50--300                                   & Inclusive                        &  $3.9\pm0.7\pm0.6$ & $6$ & SR13   \\ \hline
\multirow{6}{*}{Inclusive}       &  \multirow{4}{*}{${\geq}600$}     & \multirow{2}{*}{\x50--150}                 & ${<} 120$                          &  $14.4\pm1.2\pm1.6$ & $20$ & SR14a  \\ \cline{4-7}
&                                    &                                          & ${\geq}1 20$                      &  $0.28\pm0.14\pm0.04$ & $0$ & SR14b  \\ \cline{3-7}
&                                    & \multirow{2}{*}{150--300}                & ${<} 120$                          &  $12.1\pm1.4\pm1.6$ & $10$ & SR15a  \\ \cline{4-7}
&                                    &                                          & ${\geq}1 20$                      &  $0.40\pm0.12\pm0.05$ & $0$ & SR15b  \\ \cline{2-7}
&  \multirow{2}{*}{${\geq}60$}         & \multirow{2}{*}{${\geq}300$}         & ${<} 120$                          &  $12.1\pm1.5\pm1.9$ & $7$ & SR16a  \\ \cline{4-7}
&                                    &                                          & ${\geq}1 20$                      &  $0.70\pm0.25\pm0.11$ & $0$ & SR16b  \\ \hline
\end{tabular}
}
\end{center}
\end{table}

\begin{table}[t!]
\topcaption{Expected and observed yields in the on-Z search regions. The first uncertainty states the statistical uncertainty, while the second represents the systematic uncertainty.}
\label{tab:onZfull}
\begin{center}
\resizebox{0.9\linewidth}{!}{
\renewcommand{\arraystretch}{1.3}
\begin{tabular}{ccccccc}\hline
$\Nbjets$                           & $\HT$ [\GeVns{}]                        & $\ptmiss$ [\GeVns{}]               & $\MT$ [\GeVns{}]                     &  Expected [events]                                                      &  Observed [events]                    & SR          \\ \hline
\multirow{8}{*}{0}        &  \multirow{4}{*}{\x60-400}          & \multirow{2}{*}{70-150}                 & ${<} 120$                          &  $266\pm5\pm39$ & $241$ & SR1a  \\ \cline{4-7}
&                                    &                                          & ${\geq}1 20$                      &  $30\pm2\pm4$ & $33$ & SR1b  \\ \cline{3-7}
&                                    & \multirow{2}{*}{150--300}                & ${<} 120$                          &  $53.8\pm2.2\pm8$ & $61$ & SR2a  \\ \cline{4-7}
&                                    &                                          & ${\geq}1 20$                      &  $5.7\pm0.8\pm0.7$ & $9$ & SR2b  \\ \cline{2-7}
&  \multirow{4}{*}{400-600}         & \multirow{2}{*}{\x50--150}                 & ${<} 120$                          &  $44.6\pm1.9\pm6.5$ & $52$ & SR3a  \\ \cline{4-7}
&                                    &                                          & ${\geq}1 20$                      &  $5.1\pm0.6\pm0.7$ & $6$ & SR3b  \\ \cline{3-7}
&                                    & \multirow{2}{*}{150--300}                & ${<} 120$                          &  $16.6\pm1.3\pm2.5$ & $17$ & SR4a  \\ \cline{4-7}
&                                    &                                          & ${\geq}1 20$                      &  $1.43\pm0.33\pm0.2$ & $1$ & SR4b  \\ \hline
\multirow{4}{*}{1}        &  \multirow{2}{*}{\x60--400}          & \x70-150                                   & \multirow{4}{*}{Inclusive}      &  $116\pm4\pm15$ & $115$ & SR5  \\ \cline{3-3}\cline{5-7}
&                                    & 150--300                                  &                                  &  $21.7\pm1.2\pm2.8$ & $19$ & SR6  \\ \cline{2-3}\cline{5-7}
&  \multirow{2}{*}{400-600}         & \x50--150                                   &                                  &  $25.2\pm1.2\pm3.6$ & $25$ & SR7  \\ \cline{3-3}\cline{5-7}
&                                    & 150--300                                  &                                  &  $7.5\pm0.8\pm1$ & $9$ & SR8  \\ \hline
\multirow{4}{*}{2}        &  \multirow{2}{*}{\x60--400}          & \x50--150                                   & \multirow{4}{*}{Inclusive}      &  $47\pm1.6\pm7.4$ & $64$ & SR9  \\ \cline{3-3}\cline{5-7}
&                                    & 150-300                                  &                                  &  $7.2\pm0.8\pm1.2$ & $6$ & SR10  \\ \cline{2-3}\cline{5-7}
&  \multirow{2}{*}{400-600}         & \x50--150                                   &                                  &  $11.7\pm1\pm2.1$ & $12$ & SR11  \\ \cline{3-3}\cline{5-7}
&                                    & 150--300                                  &                                  &  $2.6\pm0.4\pm0.4$ & $6$ & SR12  \\ \hline
$ {\geq}3 $                 &  \x60--600                            & \x50--300                                   & Inclusive                        &  $4.7\pm0.5\pm0.9$ & $5$ & SR13   \\ \hline
\multirow{6}{*}{Inclusive}       &  \multirow{4}{*}{${\geq}600$}     & \multirow{2}{*}{\x50--150}                 & ${<}120$                          &  $33\pm2\pm4$ & $42$ & SR14a  \\ \cline{4-7}
&                                    &                                          & ${\geq}1 20$                      &  $4.6\pm0.6\pm0.6$ & $6$ & SR14b  \\ \cline{3-7}
&                                    & \multirow{2}{*}{150--300}                & ${<}120$                          &  $15.8\pm1.2\pm2$ & $13$ & SR15a  \\ \cline{4-7}
&                                    &                                          & ${\geq}1 20$                      &  $1.9\pm0.3\pm0.2$ & $4$ & SR15b  \\ \cline{2-7}
&  \multirow{2}{*}{${\geq}60$}         & \multirow{2}{*}{${\geq}300$}         & ${<}120$                          &  $19.1\pm1.1\pm2.8$ & $23$ & SR16a  \\ \cline{4-7}
&                                    &                                          & ${\geq}1 20$                      &  $2.28\pm0.35\pm0.26$ & $5$ & SR16b  \\ \hline
\end{tabular}
}
\end{center}
\end{table}

\begin{table}[b!]
\topcaption{Expected and observed yields in the super signal regions. The background events containing top quark(s) in association with a W, Z or Higgs boson, except \ttZ, or another pair of top quarks are denoted as $\ttbar$X. The first uncertainty states the statistical uncertainty, while the second represents the systematic uncertainty.}
\label{tab:SSRfull}
\begin{center}
\renewcommand{\arraystretch}{1.2}
\resizebox{0.9\linewidth}{!}{
\begin{tabular}{l|c|c|c|c}
 & SSR1 & SSR2  & SSR3  &  SSR4 \\ \hline
Nonprompt  &   $0.63\pm0.38\pm0.19$   & $0.00\pm0.00^{+0.3}_{-0.0}$    & $0.46\pm0.37\pm0.14$    & $0.21^{+0.23}_{-0.21}\pm0.06$  \\
$\ttZ$          &   $0.14\pm0.06\pm0.03$   & $0.05\pm0.03\pm0.01$             & $1.27\pm0.18\pm0.31$    & $0.54\pm0.10\pm0.13$               \\
$\ttbar$X    &    $0.23\pm0.04\pm0.05$   & $0.11\pm0.04\pm0.02$             & $0.50\pm0.07\pm0.08$    & $0.17\pm0.03\pm0.02$              \\
WZ             &    $0.01\pm0.01\pm0.01$   & $0.01\pm0.01\pm0.01$               & $1.03\pm0.28\pm0.21$    & $0.01\pm0.01\pm0.01$               \\
Rare          &     $0.12\pm0.06\pm0.05$   & $0.01\pm0.01\pm0.01$             & $0.40\pm0.09\pm0.14$    & $0.01\pm0.01\pm0.01$              \\ \hline
Total          &     $1.1\pm0.4\pm0.2$         & $0.18\pm0.05^{+0.3}_{-0.02}$   & $3.7\pm0.5\pm0.4$        & $0.94^{+0.26}_{-0.23}\pm0.15$   \\ \hline
Observed  &    0  & 0 & 6 & 2 \\
\end{tabular}
}
\end{center}
\end{table}

Search regions providing the best sensitivity to new physics scenarios
depend on the considered models and their parameters. In the
non-compressed scenario of the T1tttt model, the most sensitive
region is off-Z SR16b  (high \ptmiss and \MT region). When considering
the compressed scenario, the contribution from SR16b region remains
the largest, up to the most compressed cases where the SR12 off-Z
region (2 b jets, medium \ptmiss and high \HT)
starts to contribute significantly. For the T5qqqqVV model in
the non-compressed scenario, the most sensitive regions are on-Z SR16b and
SR15b
(high and medium \ptmiss, high \HT and high \MT values). When moving
towards more compressed scenarios, the most significant contributions
come from the SR16b and SR15b on-Z regions, until reaching the
compressed scenario where the most sensitive region is SR4b
(medium \ptmiss, high \HT and high \MT). The exclusion limit for
T6ttWW model is dominated by both off-Z SR16
regions (high \ptmiss region).
For the T6ttHZ model with
$\mathcal{B}(\stoptwo \to \stopone \PZ$) = 0\%,
the limits in the non-compressed scenario are driven by the off-Z SR15a (high \HT, medium \ptmiss, low \MT), while for compressed case by off-Z SR13 (high
$\Nbjets$, low and medium \HT and \ptmiss).
For $\mathcal{B}(\stoptwo \to \stopone \PZ$) = 50\% in the
non-compressed scenario, the on-Z SR16b region dominates the the
exclusion limit, while in the compressed scenario the on-Z SR13 (high \Nbjets)
and SR15b (high \HT, medium \ptmiss, high \MT) give the highest contribution. Finally, for $\mathcal{B}(\stoptwo \to \stopone \PZ$) = 100\% the on-Z SR16b plays the leading role in both compressed and non-compressed scenarios.

\clearpage

\clearpage

\section{Conclusions}
\label{sec:conclusions}

A search for physics beyond the standard model in final states with at least three electrons or muons in any combination, jets, and missing transverse momentum has been presented using data collected by the CMS detector in 2016 at $\sqrt{s} = 13\TeV$, corresponding to an integrated luminosity of 35.9\fbinv. The
analysis makes use of control regions in data to estimate reducible backgrounds and to validate simulations used to estimate irreducible background processes. To maximize sensitivity to a broad range of possible signal models, 46 exclusive signal regions are defined. No significant deviation from the expected standard model background is observed in any of these signal regions.

The results are interpreted using a simplified gluino-pair production model that features cascade decays producing four top quarks and two neutralinos. In this model, gluinos with a mass up to 1610\GeV are excluded in the case of a massless LSP. The maximum excluded LSP mass is 900\GeV.  This represents an improvement of approximately 435 and 250\GeV, respectively, compared to the exclusion limit set in a similar search based on data collected with the CMS detector in 2015, corresponding to an integrated luminosity of $2.3\fbinv$~\cite{SUS-16-003}.

For the simplified model of gluino-gluino production with decay to light-flavor quark jets, two vector bosons and neutralinos, gluino masses up to 1160\GeV and neutralino masses up to 680\GeV can be excluded.  The limit on gluino and neutralino masses extends the corresponding limit from the previous analysis by about 335 and 180\GeV, respectively.

For a simplified model of bottom squark pair production decaying to top quarks, W bosons and neutralinos, bottom squark masses up to 840\GeV are excluded for a low mass chargino, while chargino masses are excluded up to 750\GeV. These extend the previous limits by 380 \GeV for each particle.

Finally, for a simplified heavy top squark pair production model with further decays to two top quarks, Higgs or Z bosons, and neutralinos, the \stoptwo mass is excluded up to 720, 780, and 710\GeV for models with an exclusive \stoptwo $\to$ \stopone H decay, an exclusive \stoptwo $\to$ \stopone Z decay, and an equally probable mix of those two decays, while the \stopone\,mass is excluded up to 430, 540, and 450\GeV for the same branching fractions. This significantly improves the results obtained with the 8\TeV dataset~\cite{Khachatryan:2014doa}.

\begin{acknowledgments}
We congratulate our colleagues in the CERN accelerator departments for the excellent performance of the LHC and thank the technical and administrative staffs at CERN and at other CMS institutes for their contributions to the success of the CMS effort. In addition, we gratefully acknowledge the computing centres and personnel of the Worldwide LHC Computing Grid for delivering so effectively the computing infrastructure essential to our analyses. Finally, we acknowledge the enduring support for the construction and operation of the LHC and the CMS detector provided by the following funding agencies: BMWFW and FWF (Austria); FNRS and FWO (Belgium); CNPq, CAPES, FAPERJ, and FAPESP (Brazil); MES (Bulgaria); CERN; CAS, MoST, and NSFC (China); COLCIENCIAS (Colombia); MSES and CSF (Croatia); RPF (Cyprus); SENESCYT (Ecuador); MoER, ERC IUT, and ERDF (Estonia); Academy of Finland, MEC, and HIP (Finland); CEA and CNRS/IN2P3 (France); BMBF, DFG, and HGF (Germany); GSRT (Greece); OTKA and NIH (Hungary); DAE and DST (India); IPM (Iran); SFI (Ireland); INFN (Italy); MSIP and NRF (Republic of Korea); LAS (Lithuania); MOE and UM (Malaysia); BUAP, CINVESTAV, CONACYT, LNS, SEP, and UASLP-FAI (Mexico); MBIE (New Zealand); PAEC (Pakistan); MSHE and NSC (Poland); FCT (Portugal); JINR (Dubna); MON, RosAtom, RAS, RFBR and RAEP (Russia); MESTD (Serbia); SEIDI, CPAN, PCTI and FEDER (Spain); Swiss Funding Agencies (Switzerland); MST (Taipei); ThEPCenter, IPST, STAR, and NSTDA (Thailand); TUBITAK and TAEK (Turkey); NASU and SFFR (Ukraine); STFC (United Kingdom); DOE and NSF (USA).

\hyphenation{Rachada-pisek} Individuals have received support from the Marie-Curie programme and the European Research Council and Horizon 2020 Grant, contract No. 675440 (European Union); the Leventis Foundation; the A. P. Sloan Foundation; the Alexander von Humboldt Foundation; the Belgian Federal Science Policy Office; the Fonds pour la Formation \`a la Recherche dans l'Industrie et dans l'Agriculture (FRIA-Belgium); the Agentschap voor Innovatie door Wetenschap en Technologie (IWT-Belgium); the Ministry of Education, Youth and Sports (MEYS) of the Czech Republic; the Council of Science and Industrial Research, India; the HOMING PLUS programme of the Foundation for Polish Science, cofinanced from European Union, Regional Development Fund, the Mobility Plus programme of the Ministry of Science and Higher Education, the National Science Center (Poland), contracts Harmonia 2014/14/M/ST2/00428, Opus 2014/13/B/ST2/02543, 2014/15/B/ST2/03998, and 2015/19/B/ST2/02861, Sonata-bis 2012/07/E/ST2/01406; the National Priorities Research Program by Qatar National Research Fund; the Programa Severo Ochoa del Principado de Asturias; the Thalis and Aristeia programmes cofinanced by EU-ESF and the Greek NSRF; the Rachadapisek Sompot Fund for Postdoctoral Fellowship, Chulalongkorn University and the Chulalongkorn Academic into Its 2nd Century Project Advancement Project (Thailand); the Welch Foundation, contract C-1845; and the Weston Havens Foundation (USA).

\end{acknowledgments}

\bibliography{auto_generated}
\cleardoublepage \appendix\section{The CMS Collaboration \label{app:collab}}\begin{sloppypar}\hyphenpenalty=5000\widowpenalty=500\clubpenalty=5000\textbf{Yerevan Physics Institute,  Yerevan,  Armenia}\\*[0pt]
A.M.~Sirunyan, A.~Tumasyan
\vskip\cmsinstskip
\textbf{Institut f\"{u}r Hochenergiephysik,  Wien,  Austria}\\*[0pt]
W.~Adam, F.~Ambrogi, E.~Asilar, T.~Bergauer, J.~Brandstetter, E.~Brondolin, M.~Dragicevic, J.~Er\"{o}, M.~Flechl, M.~Friedl, R.~Fr\"{u}hwirth\cmsAuthorMark{1}, V.M.~Ghete, J.~Grossmann, J.~Hrubec, M.~Jeitler\cmsAuthorMark{1}, A.~K\"{o}nig, N.~Krammer, I.~Kr\"{a}tschmer, D.~Liko, T.~Madlener, I.~Mikulec, E.~Pree, D.~Rabady, N.~Rad, H.~Rohringer, J.~Schieck\cmsAuthorMark{1}, R.~Sch\"{o}fbeck, M.~Spanring, D.~Spitzbart, W.~Waltenberger, J.~Wittmann, C.-E.~Wulz\cmsAuthorMark{1}, M.~Zarucki
% ...
\vskip\cmsinstskip
% ...
1:~~Also at Vienna University of Technology, Vienna, Austria\\
% ...

\end{sloppypar}
\end{document}